\documentstyle[12pt,psfig]{article}

\begin{document}

\title{Translocation Properties of Primitive Molecular Machines and
  Their Relevance to the Structure of the Genetic Code}
\author{M Aldana-Gonz\'alez$^\dag$, G Cocho$^\S$, \\
  H. Larralde$^\P$, G Mart\'\i nez-Mekler$^\P$.}

\date{} 

\maketitle

\begin{center}
  \small \it $^\dag$The James Franck Institute, The University of
  Chicago 5640 South Ellis Avenue, Chicago, Il, 60637, US.\\
  $^\S$Instituto de F\'{\i}sica, UNAM. Apdo. Postal 20-364, 01000
  M\'exico D.F.,  M\'exico.\\
  $^\P$Centro de Ciencias F\'{\i}sicas, UNAM. Apdo. Postal 48-3, 62251
  Cuernavaca, Morelos, M\'exico.
\end{center}

\begin{center}
\it Submitted to the Journal of Theoretical Biology.\\
 November 2001.
\end{center}

\begin{abstract}
  We address the question, related with the origin of the genetic
  code, of why are there three bases per codon in the translation to
  protein process. As a followup to our previous work,
  \cite{max1,max2,max3} we approach this problem by considering the
  translocation properties of primitive molecular machines, which
  capture basic features of ribosomal/messenger RNA interactions,
  while operating under prebiotic conditions.  Our model consists of 
  a short one-dimensional chain of charged particles(rRNA antecedent)
  interacting with a polymer (mRNA antecedent) via electrostatic
  forces. The chain is subject to external forcing that causes it to
  move along the polymer which is fixed in a quasi one dimensional
  geometry. Our numerical and analytic studies of statistical
  properties of random chain/polymer potentials suggest that, under
  very general conditions, a dynamics is attained in which the chain
  moves along the polymer in steps of three monomers. By adjusting the
  model in order to consider present day genetic sequences, we show
  that the above property is enhanced for coding regions.  Intergenic
  sequences display a behavior closer to the random situation. We
  argue that this dynamical property could be one of the underlying
  causes for the three base codon structure of the genetic code
\end{abstract}


\section{Introduction}

The origin of the genetic code has been a subject of intense research
since its structure was completely elucidated in the early 1970's.  In
subsequent years, the scientific community has produced several
theories in order to explain why the genetic code has this structure.
Among these theories, the most prominent ones are the stereochemical
theory, the frozen-accident theory and the coevolutionary theory
\cite{alberti1}-\cite{freeland}. Roughly speaking, these theories try
to account for the structure of the genetic code by looking at the
interactions between codons and amino acids, the biosynthetic
relationships among different amino acids and how the metabolic
pathways between them have been selected throughout evolution.
Nevertheless, the fact that all the codons are made up of three
nucleotides, has mostly been taken for granted and barely brought into
question.

One of the most widely used arguments found in the literature to
explain the trinucleotide codon structure of the genetic code, was
given by Sidney Brenner in 1961 \cite{brenner1,brenner2}. According to
this argument, codons are made up of three nucleotides (or bases, for
short) because there are 20 amino acids to be specified by the genetic
information expressed by a 4 letter ``alphabet'' (the four bases
A,G,C,U).  If codons were composed of only two bases, there would be
only 16 different combinations ($4^2$), which are not enough to
specify for 20 amino acids.  If instead, codons were made up of more
than three bases, there would be at least 256 combinations ($4^4$),
and these are too many for only 20 amino acids. Hence, less than three
bases per codon are not enough, and more than three would imply an
excessive degeneration of the code.  The result coming out from this
argument is that three bases per codon is the optimal ``bit of
information'' that can be used in order to specify for the 20
different amino acids by means of a 4 letter ``alphabet''.

The above argument, however, does not constitute an explanation by
itself, mainly because it only moves the question of ``why three?'' to
the questions of ``why twenty?'' or for that matter ``why four?''.
There is no reason for the genetic information to codify for only 20
amino acids since living organisms use more than those specified by
the genetic code \cite{lehninger}. In addition, this argument assumes
that all the codons must have the same length (number of bases), even
though more efficient codes can be obtained by allowing the length of
the codons to vary \cite{doig}. Finally, \emph{given} that 20 amino
acids have to be specified by using 4 different bases, Brenner's
argument leads to the simplest code that might be thought of. But even
in such a case, simplicity has to be accounted for as a relevant
criterium.

In this work we address the question of the origin of the three-base
codon structure of the genetic code from a dynamical point of view. We
consider a simple molecular machine model which captures some of the
principal features of the interaction between primitive realizations
of the ribosome and of the mRNA.  Our main objective is to present a
dynamical scenario, compatible with prebiotic conditions, of how the
triplet structure of the genetic code could have arisen.  The model we
propose is a follow up of the one introduced by Aldana, Cocho and
Mart\'{\i}nez-Mekler \cite{max1,max2,max3} and is consistent with the
current evidence suggesting the ``RNA world'' hypothesi1s
\cite{rnaworld}. In this scheme the crucial molecules involved in the
prebiotic and protobiotic processes, that eventually led to
codification and translation mechanisms of the genetic information,
were RNA related.q

\begin{figure}[h]
\psfig{file=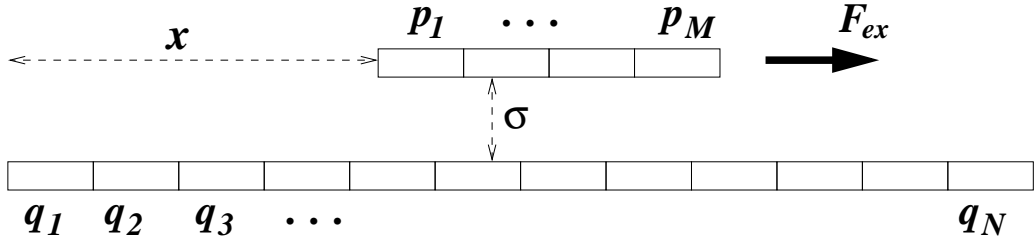,width=\hsize,clip=}
\caption[]{The molecular machine model. A one dimensional chain composed 
  of $M$ monomers interacts with a one dimensional polymer, $N$
  monomers in length.  The charges $\{p_i\}$ along the chain, as well
  as the charges $\{q_j\}$ of the polymer are independent random
  variables.  The chain is at a fixed perpendicular distance $\sigma$
  from the polymer, but can move along the polymer. $x$ is the
  horizontal position of the chain with respect to the polymer.  }
\label{fig1}
\end{figure}

In our model, based on the setup depicted in Fig. 1, a short
one-dimensional polymer composed of $M$ monomers interacts with a much
longer one, via electrostatic forces. In order to avoid confusion,
from now on we will refer to the short polymer as ``the chain'', and
to the long polymer simply as ``the polymer''. The electrostatic
interaction between the chain and the polymer is due to the presence
of electric charges, or multipolar moments, in the monomers of both
the chain and the polymer.

The charges of the monomers of the chain and of the polymer are
assigned at random following a uniform distribution. Therefore, the
resulting chain-polymer interaction potential has a random profile.
The chain is allowed to move along the polymer, but is constrained to
remain at a fixed perpendicular distance $\sigma$ from it.
Consequently, transport is one-dimensional. One of our main results is
to show that under very general conditions, a dynamics is attained in
which the chain moves along the polymer in effective ``steps'' whose
mean length is three monomers.  We argue that this dynamical feature
may be one of the underlying causes of the three base codon structure
of the genetic code.

This paper is organized as follows: section \ref{themodel} describes
in detail the model and the assumptions introduced. In section
\ref{unichain} we recall some statistical aspects of our previous
analysis \cite{max1, max2, max3} of the random interaction potentials
between the chain and the polymer for the simplest case in which the
former is composed of just one particle ($M=1$). We exhibit
numerically that, even in this simple case, the mean distance
$\bar{d}$ between consecutive minima along the interaction potential
is very close to three: $\bar{d}\sim 3$ (taking the monomer length as
spatial unit). After retrieving the analytical expression for this
distance \cite{max1}, we then look into the probability ${\mathbf
  P}_m(d)$ of two neighboring potential minima being separated by a
distance $d$. Subindex $m$ refers to the number of different types of
monomers in the polymer and in the chain.  This probability function
shows that, even though the mean distance $\bar{d}$ is close to three,
the most probable distance between consecutive minima is $d^*=2$ for
$m>2$.  In section \ref{realseq} the monomer charges along the polymer
are assigned in correspondence with protein-coding regions of the
genome of real organisms (e.g. \emph{Drosophila} or \emph{E.  coli})
instead of at random. For this case, the probability function
${\mathbf P}_m(d)$ is modified so that not only the mean distance is
$\bar{d}\sim 3$, but also the most probable one happens to be $d^*=3$.

In section \ref{polychain} we introduce the more realistic case $M>1$,
which takes into account the fact that the ribosome is not a point
particle, that it has spatial structure and presents several
simultaneous contact points between its own rRNA and the mRNA polymer.
For small chain lengths ($M\sim 10$), the probability distribution
${\mathbf P}_m(d)$ is indicative of wide fluctuations and has a form
strongly dependent on the particular assignment of charges in the
chain. One of our main findings is that for such chains the most
likely configurations are those in which both, the mean distance
$\bar{d}$ and the most probable one $d^*$ are equal to three,
\emph{even when the monomer charges along the polymer and the chain
  are assigned at random}. In section \ref{dynamics} we analyze the
dynamics resulting from the model when an external force is pulling
the chain, forcing it to move as a rigid object along the polymer. The
power spectrum of the velocity of the chain reveals that, under some
very general circumstances, for small chain lengths ($M\sim 10$), there
is a sharp periodicity in the dynamics of the system, with a slowing
down of the velocity of the chain every three monomers. Finally,
section \ref{summary} is devoted to the discussion of the results and
their relevance to the origin of the genetic code.


\section{The model}
\label{themodel}

The model we propose consists of a chain of $M$ monomers interacting
with a very long polymer composed of $N$ monomers, with
$N\rightarrow\infty$ (see Fig.1). The chain is constrained to remain
at a given distance $\sigma$ perpendicular to the polymer and is
allowed to move in along the polymer, we shall define $x$ as its
position in this direction relative to the polymer. We will denote the
monomer charges in the chain and in the polymer by $\{p_i\}$ and
$\{q_j\}$, respectively. We should mention that by ``charge'' we do
not necessarily mean Coulomb charge.  Both $\{p_i\}$ and $\{q_j\}$
could be dipolar moments, induced polarizabilities, or similar
quantities resulting from electrostatic interactions between chain
monomers and polymer monomers with potentials of the form
$1/r^\alpha$, where $\alpha$ characterizes the ``charge'' type.

We will assume that all the monomers in the chain, and separately,
all the monomers in the polymer, are of the same nature, namely,
all of them are either Coulomb charges, or dipolar moments, or
polarizable molecules, etc.  In addition, taking into account that
in the origin of life conditions the genetic molecules were not
yet likely to convey any structured information, we will consider
the charges $\{p_i\}$ and $\{q_j\}$ to be discrete independent
random variables, acquiring one of the $m$ different values
$\xi_1,\ \xi_2,\ \dots,\ \xi_m$ with the same probability. Hence,
the probability function $P(q)$ for both, the $\{p_i\}$ and
$\{q_j\}$ variables, will be

\begin{equation}
P(q)=\frac{1}{m}\sum_{j=1}^{m}\delta(q-\xi_j)
\end{equation}
where $\delta(q)$ is the Dirac delta function.  In general, in
this work we will take the values $\xi_1,\ \xi_2,\ \dots,\ \xi_m$
as integers.  Parameter $m$ represents the number of different
types of monomers from which the polymer and the chain are made
of.  For the case of real genetic sequences $m=4$, but we will not
restrict the value of $m$ to be 4.

All the monomers will have the same length $L$, which we take as
the spatial unit of measure: $L=1$. We also assume the charge in
each of the monomers to be uniformly distributed along the length
$L$, so that the charge density $\lambda_j(x)$ in the jth-monomer
of the polymer, for example, is a constant whose value is
$\lambda_j(x)=q_j/L$. Nevertheless, it is worth mentioning that
the dynamics of the model does not depend strongly on the
particular shape of the monomer charge density $\lambda_j(x)$, as
long it is a smooth function of $x$ (``smooth'' in the sense of
differentiability).

With the preceding assumptions, the interaction potential $V_{ij}(x)$
between the ith-monomer in the chain and the jth-monomer in the
polymer is given by

\begin{equation}
V_{ij}(x)=Kp_i q_j\int_{x+i-1}^{x+i}\int_{j-1}^{j}\frac{dx' dx''}
{\left[(x'-x'')^2 + \sigma^2\right]^{\alpha/2}}
\label{Vmm}
\end{equation}
where $K$ is a constant whose value depends on the unit system
used to measure the physical quantities. In the above expression,
$L$ has already been set equal to $1$. Parameter $\alpha$
characterizes the kind of interaction between the chain and the
polymer: $\alpha=1$ corresponds to an ion-ion interaction,
$\alpha=2$ represents an ion-dipole interaction, and so forth.
Note that this parameter does not depend on the indices $i$ and
$j$, since all the monomers in the chain are of the same nature
and those in the polymer are themselves of the same nature,
differing from each other only in the value of the charge they
contain.  The overall interaction potential $V_\sigma^\alpha(x)$
between the whole chain and the entire polymer is given by the
superposition of the individual potentials $V_{ij}(x)$:

\begin{equation}
V_\sigma^\alpha(x)=\sum_{i=1}^{M}\sum_{j=1}^{N}V_{ij}(x)
\label{Vcp}
\end{equation}

Equations (\ref{Vmm}) and (\ref{Vcp}) establish the type of random
potentials we will be considering. Our first aim is to analyze the
spatial structure of these potentials, giving their statistical
characterization. This will be done in the three following
sections. Subsequently, we will consider the dynamics of the chain
moving along the polymer interacting with it by means of a random
potential, subject to an external driving force and seek under
what conditions, if any, transport in ``steps'' of three monomers
can be achieved.


\section{Random potentials: $M=1$}
\label{unichain}

Let us start with the simplest case $M=1$, in which the chain consists
of just one monomer.  We will refer to this case as the
``single-monomer-chain'' case, and to the chain simply as ``the
particle''. The reason to consider this simple situation is twofold:
on one hand, it is useful in order to introduce the relevant ideas
behind the model. On the other hand, it is simple enough as to obtain
exact analytical results in a more o less straightforward way.  In
previous work we have already analyzed some statistical properties of
the random potentials given by expressions (\ref{Vmm}) and (\ref{Vcp})
for the case $M=1$ \cite{max1,max2,max3}. After a short review of some
of those results we center our attention on the probability
distribution ${\mathbf P}_m(d)$.

The overall particle-polymer interaction potential is given by

\begin{equation}
V_\sigma^\alpha(x)=\sum_{j=1}^{N}q_j\int_{x}^{x+1}\int_{j-1}^{j}
\frac{dx' dx''}{\left[(x'-x'')^2 + \sigma^2\right]^{\alpha/2}}
\end{equation}

Note that in the previous expression we have set $Kp_1$, the only
charge in the chain, equal to one. Fig.2 shows three graphs of the
potential $V_\sigma^\alpha(x)$ for $\sigma=0.5$ and different values
of the parameter $\alpha$. To generate these graphs, the following
probability function for the charges $\{q_j\}$ was used:

\begin{equation}
P(q)=\frac{1}{6}\sum_{\mu=-3}^{3}\delta(q-\mu)
\end{equation}
namely, each one of the variables $\{q_j\}$ acquired one of the six
different values $\{\pm 1, \pm 2, \pm 3\}$ with probability $1/6$
($m=6$). In Fig.2, the random realization of the charges $\{q_j\}$
along the polymer was the same for the three graphs. As can be seen
from this figure, the distribution of maxima and minima along the
potential does not change by varying the value of the parameter
$\alpha$, in the sense that all the maxima and minima remain
essentially at the same positions. What occurs as $\alpha$ takes
larger and larger values is that the potential becomes a step-like
function.

\begin{figure}
\psfig{file=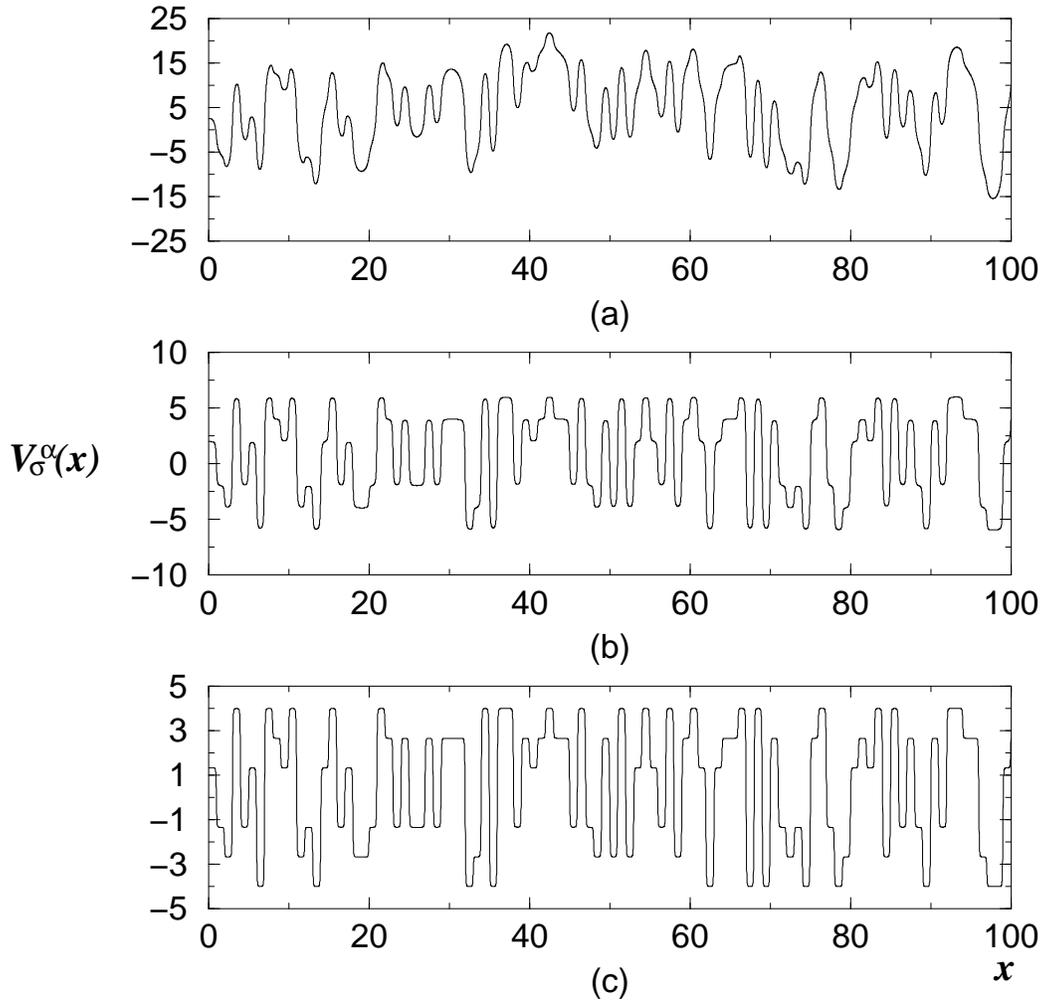,width=\hsize,clip=}
\caption[]{ Particle-polymer interaction potential for $\sigma=0.2$
  and different values of $\alpha$. (a) $\alpha=1$; (b) $\alpha=3$;
  (c) $\alpha=5$. The charges along the polymer, selected at random,
  are the same in the three graphs shown. Note that as $\alpha$
  increases, the potential becomes a step-like function, but the
  positions of the maxima and minima do not change.  Also note that
  there are 33 potential minima distributed along 100 monomers.
  Therefore the mean distance between consecutive minima is
  $\bar{d}=100/33=3.03$.  }
\label{fig2}
\end{figure}

Fig.3 presents an analogous situation, but now keeping $\alpha$
constant ($\alpha=2$) and varying $\sigma$. The behavior of the
potential is similar to the previous case: the potential becomes a
step-like function as $\sigma$ decreases and the positions of the
maxima and minima are not appreciably modified.

The above considerations exhibit that for small values of $\sigma$,
say $0<\sigma\leq 1$, the distribution of maxima and minima along the
potential is entirely determined by the distribution of charges along
the polymer and is independent of the particular values acquired by
the parameters $\alpha$ and $\sigma$. Therefore, in order to find out
the distribution of maxima and minima along the interaction potential,
it is possible to substitute the continuous random potential given by
expressions (\ref{Vmm}) and (\ref{Vcp}), by the equivalent step-like
potential defined by

\begin{equation}
V(x)=\sum_{j=1}^{N}V_j\left[H\left(x-(j-1)\right)-H(x-j)\right]
\label{Vesc}
\end{equation}
where $H(x)$ is the Heaviside function\footnote{The Heaviside function
  $H(x)$ is defined as $H(x)=0$ if $x<0$ and $H(x)=1$ if $x\geq 0$},
and $V_j$ is a random variable whose value is directly proportional to
the charge $q_j$ of the jth-monomer in the polymer:

\begin{equation}
V_j=K p_1 q_j
\label{Vparticle}
\end{equation}
where $p_1$ is the charge of the particle (the only monomer in the
chain).

\begin{figure}
\psfig{file=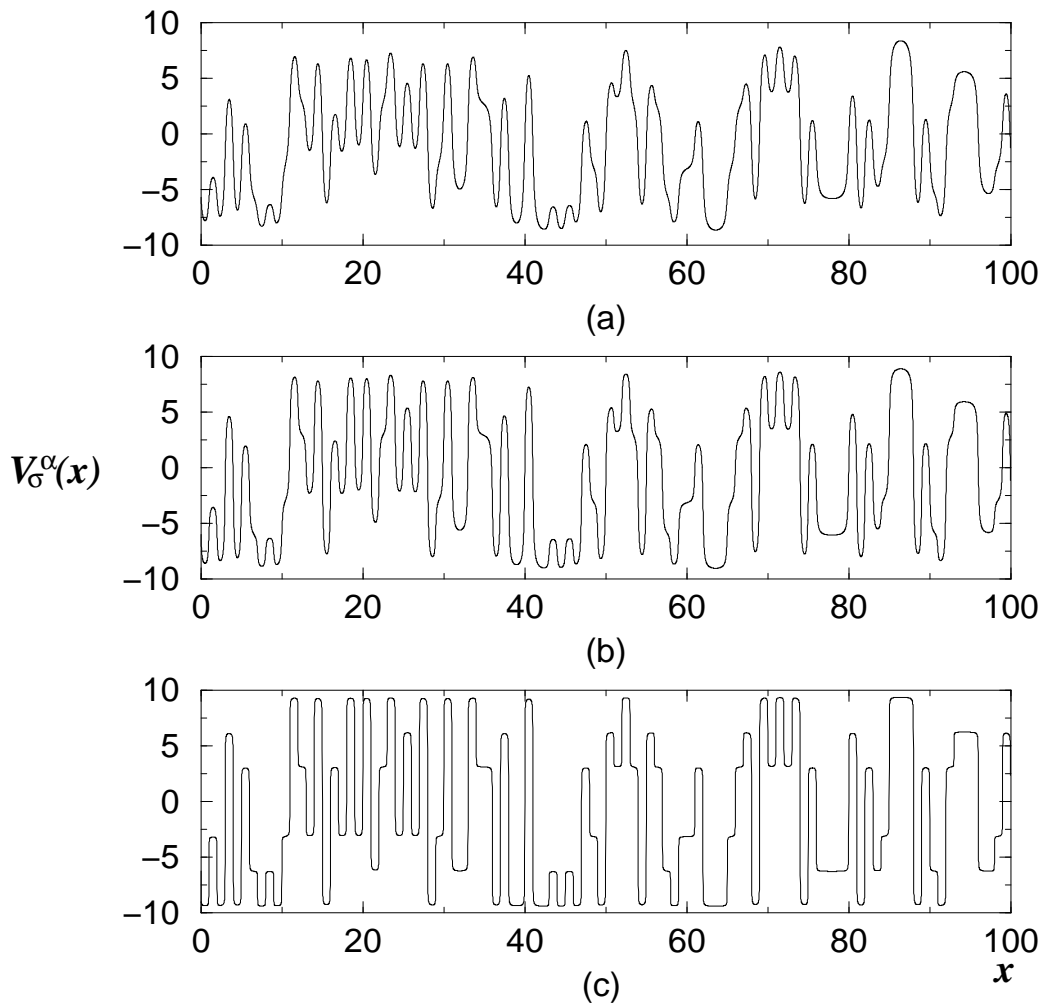,width=\hsize,clip=}
\caption[]{Particle-polymer interaction potential for $\alpha=2$ and
  different values of $\sigma$. (a) $\sigma=0.2$; (b) $\sigma=0.1$;
  (c) $\sigma=0.01$. As $\sigma$ acquires smaller values, the
  interaction potential becomes a step-like function. As before, the
  distribution of maxima and minima does not change when
  $\sigma\rightarrow 0$. Note that in this case there are 34 minima in
  the interval $[0,100]$. Therefore, the mean distance between
  neighboring minima is $\bar{d}=100/34=2.94$.  }
\label{fig3}
\end{figure}

Since the random charges $\{q_j\}$ are statistically independent, so
are the $\{V_j\}$. Expression (\ref{Vesc}), which we will refer to as
the {\bf step-like limit}, is suitable for the analytical
determination of the probability function of the distances between
consecutive potential minima. This probability function gives
important information concerning the dynamics of the system.  If some
external force is acting on the particle (or the chain), forcing it to
move in one direction (right or left) along the polymer, the particle
will spend more time in the energy minima than in the maxima. Such a
movement may be interpreted by considering the particle as ``jumping''
from one minimum to the next (see Fig.4).  It is worth noticing that
the mean distance between consecutive minima in the potentials shown
in Fig.2 and Fig.3 is nearly three. In Fig.2 there are 33 minima
distributed among 100 monomers, and consequently the mean distance
between consecutive minima in this case is $\bar{d}=100/33 \simeq
3.03$.  Analogously, the mean distance between neighboring minima in
Fig.3 is $\bar{d}=100/34\simeq 2.94$. Therefore, it is expected that
in its motion along the polymer, the velocity of the particle will
slow down, on average, every three monomers, being momentarily
``trapped'' in each of the potential minima.

\begin{figure}
\psfig{file=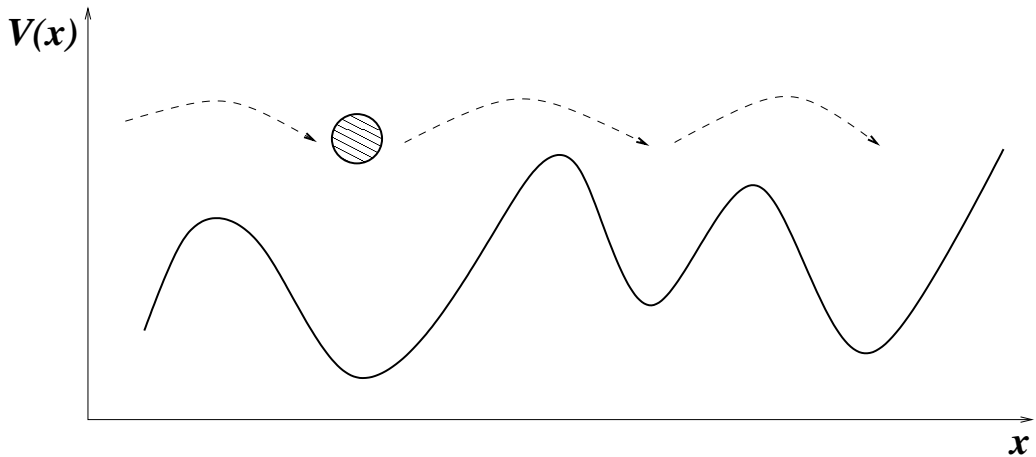,width=\hsize,clip=}
\caption[]{ In a one dimensional potential, the dynamics is
  determined by the distribution of maxima and minima. If a molecule
  is moving along the potential and is subject to an external force,
  it will spend more time in the potential minima than in the maxima.
  This kind of motion can be considered as if the molecule were
  ``jumping'' from one minimum to the next.}
\label{fig4}
\end{figure}

\begin{figure}
\psfig{file=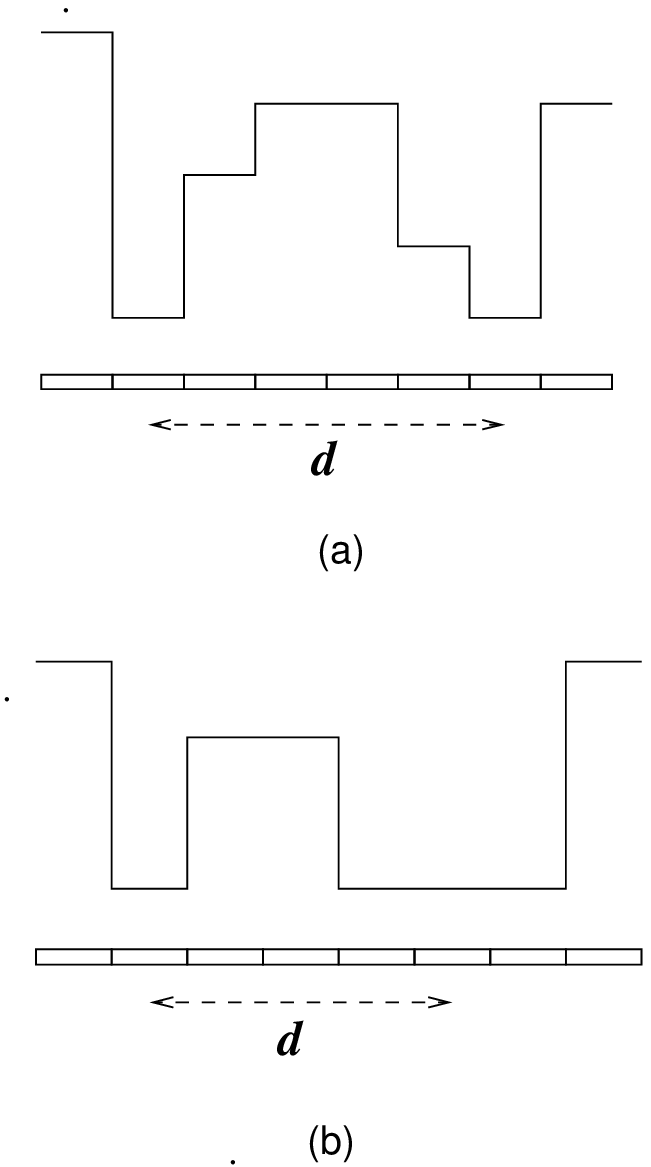,width=2.6in,clip=}
\caption[]{ In the step-like limit, the probability function
  ${\mathbf P}_m(d)$ can be calculated by counting all the possible
  configurations like the one shown in this figure. (a) Two minima,
  one at $V_j$ and the other at $V_{j+d}$ are separated by a distance
  $d$, with no other minima in between. (b) The extended minimum on
  the right is due to the fact that several adjacent monomers acquired
  the same charge value. The distance $d$ will be measured between the
  mid points of the two minima. }
\label{fig5}
\end{figure}

By using the step-like limit, in reference \cite{max1} we have
shown that the mean distance $\bar{d}$ between consecutive
potential minima for a long polymer (the large N limit) is given
by

\begin{equation}
\bar{d}=\frac{6m}{2m-1}
\label{germi}
\end{equation}
where $m$ is the number of different monomer types. The above
equation shows that the mean distance $\bar{d}$ is always between
3 and 4, and approaches 3 asymptotically as $m\rightarrow\infty$.
In particular, for $m=4$ (the biological value) we have
$\bar{d}\simeq 3.43$.

In order to characterize the fluctuations around the mean distance
$\bar{d}$, it is useful to compute the probability distribution
function ${\mathbf P}_m(d)$, which we recall, gives the probability
of two consecutive minima being separated by a distance $d$ when
there are $m$ different types of monomers. In the step-like limit,
this computation is carried out by counting all the configurations
of the step-like variables $\{V_j\}$ in which there are two
minima, one at $V_j$ and the other at $V_{j+d}$, with no other
minima in between. The situation is illustrated in Fig.5a.  Since
in the step-like limit the interaction potential is constant along
every monomer, we will adopt the convention to measure the
distance $d$ between two adjacent minima from the mid point of the
first minimum to the mid point of the second one, as illustrated
in Fig.5b. With the above convention, the resulting distances $d$
can only acquire integer or half-integer values. For a finite
number $m$ of different charges, the explicit calculation of the
probability function ${\mathbf P}_m(d)$ consists mainly on
counting configurations, though conceptually straightforward, it
involves a considerable amount of algebra. Here we present the
final expressions:

\begin{equation}
{\mathbf P}_m(d)=\frac{6}{2m-1}\sum_{k=0}^{d-2}
\frac{2k+1}{m^{k+d+2}}N_m(d-k)
\end{equation}
if $d$ is integer: $d = 2,\ 3,\ 4,\dots$, and
\begin{equation}
{\mathbf P}_m(d)=\frac{12}{2m-1}\sum_{k=0}^{d'-2}
\frac{2k+1}{m^{k+d'+3}}N_m(d'-k)
\end{equation}
if $d$ is half-integer: $d=5/2,\ d=7/2,\ d=9/2,\dots$, where
$d'=Int[d]$. In the above expressions, $N_m(d)$ is a polynomial
whose degree and coefficients depend on $m$.  For $m=2$, $m=3$ and
$m=4$ the polynomials are given by

\begin{eqnarray}
N_2(d)&=&1\nonumber\\
N_3(d)&=&1+2d+2d^2\\
N_4(d)&=&-2+\frac{1}{8}\left[30d+73d^2+22d^3+3d^4\right]\nonumber
\end{eqnarray}

For the case of $m=\infty$ we have also derived a closed expression
\cite{hernan}, which has a much simpler form:

\begin{equation}
{\mathbf P}_{\infty}(d)=3\frac{2^d}{(d+3)!}\left(d^2+d-2\right)
\label{minfty}
\end{equation}

\begin{figure}
\psfig{file=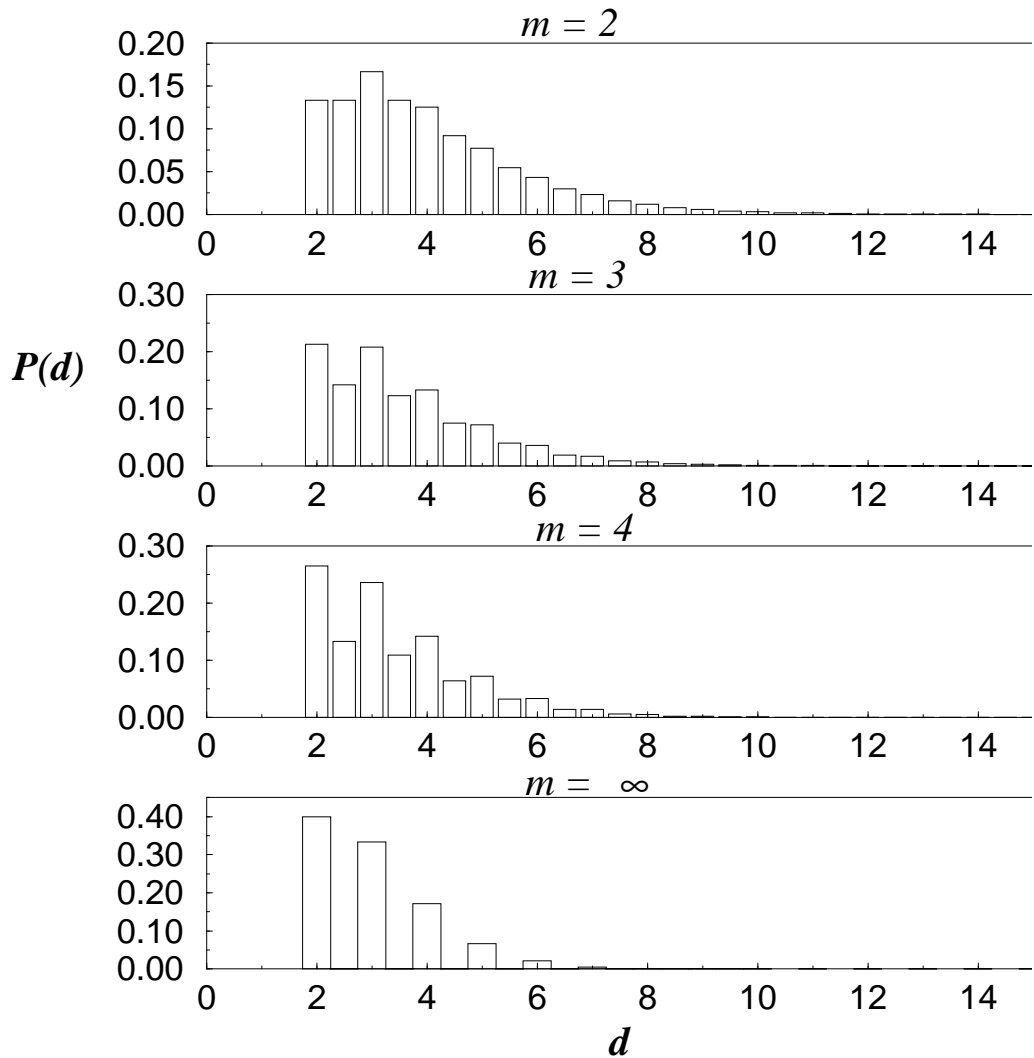,width=\hsize,clip=}
\caption[]{ Probability function ${\mathbf P}_m(d)$ for different
  values of $m$. These graphs correspond to the case in which the
  charges along the polymer are assigned at random. Note that for
  $m>2$, the most probable distance $d^*$ occurs at $d^*=2$. }
\label{fig6}
\end{figure}

The preceding distributions are plotted in Fig.6. It can be seen from
this figure that the most probable distance $d^*$ between consecutive
potential minima is $d^*=2$, except for the case $m=2$ in which
$d^*=3$. Hence, according to the transport mechanism suggested in
Fig.4, whenever there are more than two different types of monomers,
the particle will move along the polymer in ``jumps'' whose mean
length is close to three, but whose most probable length is actually
two. The difference between the mean distance $\bar{d}$ and the most
probable distance $d^*$ is due to the presence of ``tails'' in the
probability function ${\mathbf P}_m(d)$. Namely, to the fact that
${\mathbf P}_m(d)$ has non zero values even for large $d$.
Nevertheless, in section \ref{polychain} we will show that these
``tails'' can be shrunk almost to zero when the chain is made up of
more than one particle ($M>1$). This is one of the main results of
this paper.

To end this section, it is worth mentioning that half-integer
distances between two neighboring minima occur when one or both of
these minima extend over several monomers (see Fig.5b). In these
configurations, the charges of the adjacent monomers constituting the
extended minimum have the same value. Configurations in which groups
of adjacent equally charged monomers occur, are less likely than
configurations in which adjacent monomers have different charges, and
the former tend to disappear as $m$ increases (see Fig.6).


\section{Real genetic sequences: $M=1$}
\label{realseq}

The charges $\{q_j\}$ along the polymer can be assigned in
correspondence with the genetic sequence of an organism, rather than
in a random way. The purpose of doing so is to find out how the
potential minima and maxima along real genetic sequences are
distributed, and to compare the resulting distribution with the one
corresponding to the random case.  Since genetic sequences are made
out of four different bases (A, U, C and G) we consider four different
possible values $\xi_1,\ \xi_2,\ \xi_3,\ \xi_4$ for the charges
$\{q_j\}$, i.e. $m=4$. To proceed further, it is necessary to
establish a correspondence between the charge values
$\xi_1,\dots,\xi_4$ and the four bases A, U, C and G. An arbitrary
look up table is the following:

\begin{equation}
\begin{array}{c|c}
\mbox{base}&\mbox{charge value}\\ \hline
\mbox{A}&-2\\
\mbox{U}&-1\\
\mbox{C}&+1\\
\mbox{G}&+2
\end{array}
\label{corres}
\end{equation}

With the above correspondence, if the jth-base in a given genetic
sequence happens to be A, for example, then the charge of the
corresponding jth-monomer in the polymer will be $q_j=-2$.

\begin{figure}
\psfig{file=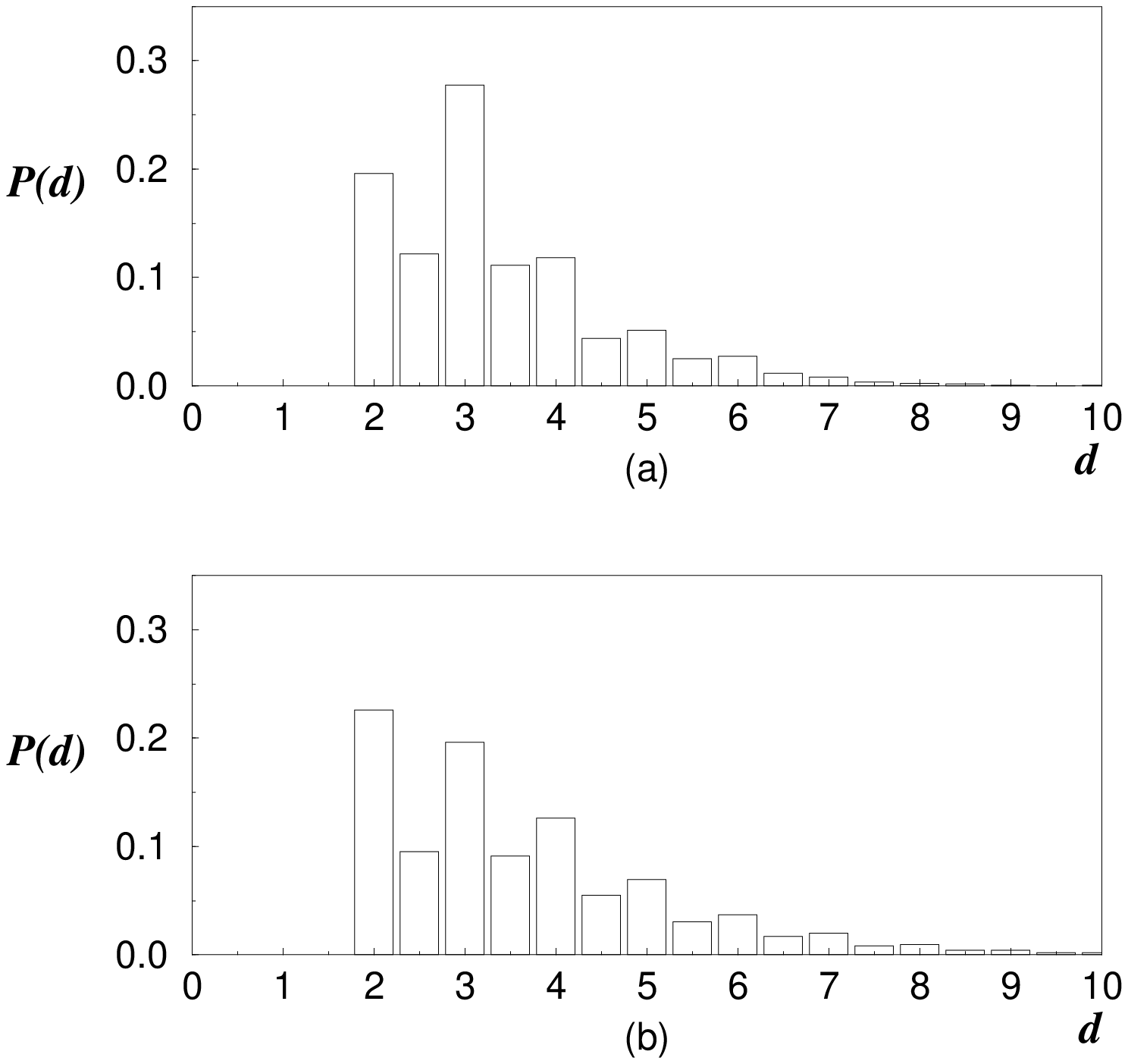,width=\hsize,clip=}
\caption[]{ Probability function ${\mathbf P}_4(d)$ computed by
  assigning the charges in the polymer in correspondence with genetic
  sequences of the \emph{Drosophila malanogaster} genome. (a) The
  genetic sequence used is a concatenation of several protein-coding
  genes. For this sequence, the mean distance between consecutive
  minima in the potential is $\bar{d}\simeq 3.15$. Note the probability
  function has its highest value at $d^*=3$. (b) In this case the sequence
  is an intergenic region of the genome (non-coding) with mean distance
  $\bar{d}\simeq 3.43$ and most probable value $d^*=2$.}
\label{fig7}
\end{figure}

Figure Fig.7a shows the probability distribution ${\mathbf P}_4(d)$
computed numerically by using a \emph{Drosophila melanogaster}
protein-coding sequence, 45500 bases in length (several genes were
concatenated to construct this sequence). The mean distance between
consecutive potential minima for this sequence is $\bar{d}\simeq 3.15$
and, as follows from the figure, the most probable distance is
$d^*=3$.  Therefore, in the ``real sequence case'' not only is the
mean distance $\bar{d}$ very close to 3, but also the most probable
distance $d^*$ turns out to be 3.  A comparison of Fig.6c with Fig.7a,
shows that for protein-coding genetic sequences, the potential minima
along the polymer are more often separated by three monomers than in
the random case. When protein-coding sequences are used, the value of
${\mathbf P}_4(d)$ increases at $d=3$ and decreases at $d=2$.

The above behavior does not occur when non-coding sequences of the
genome are used for the monomer charge assignment along the polymer.
Fig.7b shows the probability function ${\mathbf P}_4(d)$ for the case
in which the monomer charges are in correspondence with an intergenic
sequence of the \emph{Drosophila's} genome. The length of the sequence
is again 45500 bases. As can be seen from the figure, in this case the
probability function ${\mathbf P}_4(d)$ looks much more like the one
obtained in the random case.

It is important to remark that the behavior of ${\mathbf P}_4(d)$
exhibited in Fig.7a for real protein-coding sequences does not depend
on the particular correspondence (\ref{corres}) between bases and
charge values being used, as long as they are of similar order of
magnitude and they can allow for an order relation. These conditions
hold for the four bases A, T, C and G which have charges of the same
order of magnitude \cite{dipolos}\footnote{The real physical charge
  values will depend on the system of units, which is contained
  basically in the constant $K$ appearing in expression
  (\ref{Vmm})}and are fulfilled by our present choice of
$\{\pm1,\pm2\}$. The order relation is necessary for the interaction
potential to have maxima and minima.  It is worth asking the effect
that changes in the order relation have on the probability function
${\mathbf P}_4(d)$. There are 24 possible order relations among the
four bases A, T, C, and G ($4!$ permutations). The 24 probability
functions ${\mathbf P}_4(d)$ corresponding to these permutations are
plotted in Fig.8a for \emph{Drosophila's} protein-coding sequence. As
can be seen from the figure, the probability functions basically
overlap, independently of the particular order relation between the
bases, with a peak at $d^*=3$. The invariance of ${\mathbf P}_4(d)$
under base permutations also holds for non-coding sequences as is
shown in Fig. 8.b., where $d^*=2$ for intergenic sequences of
Drosophila \emph{Drosophila's}. This value of $d^*$ suggests that
non-coding sequences behave as random structures.

\begin{figure}
\psfig{file=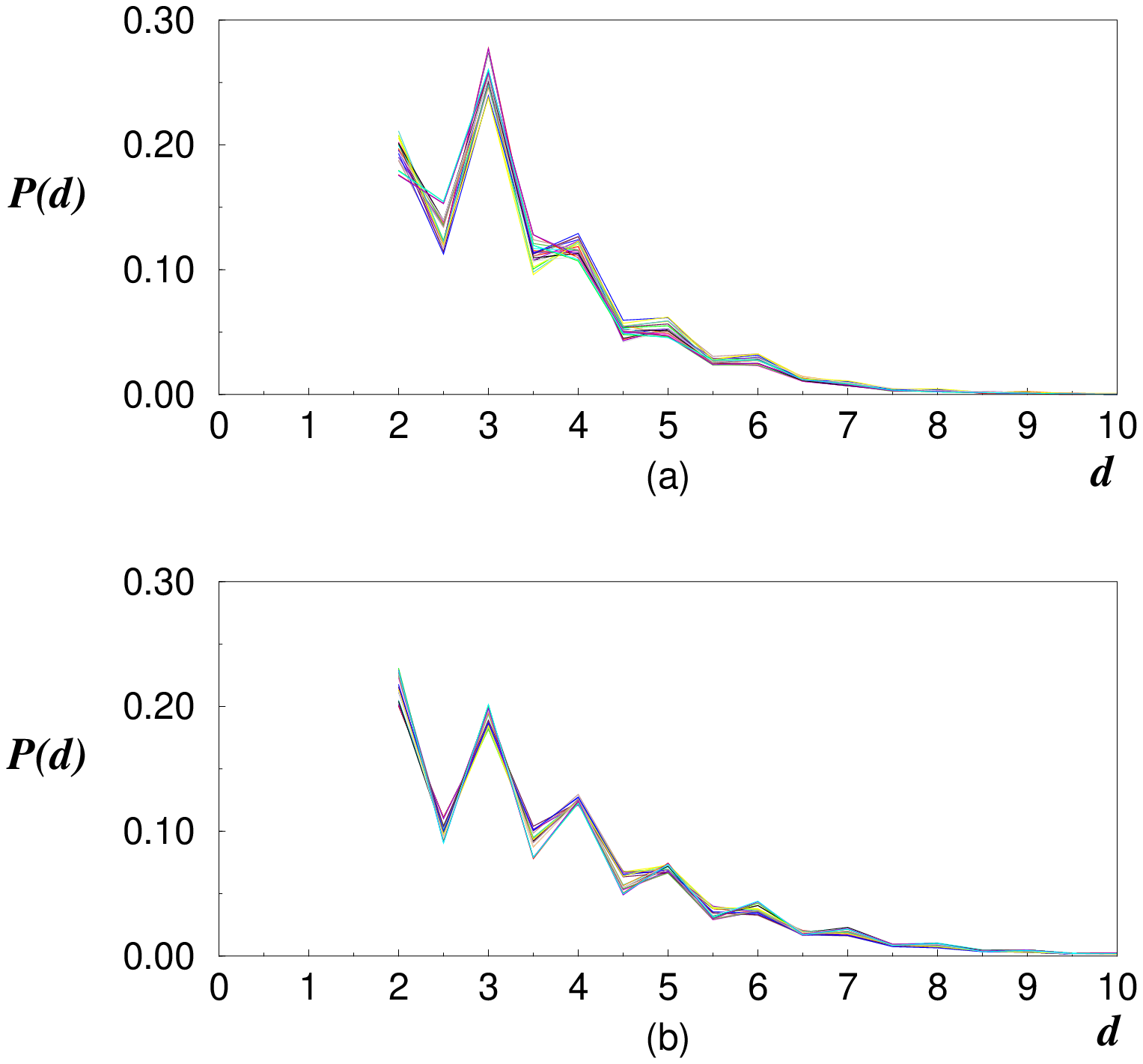,width=\hsize,clip=}
\caption[]{ Plot of the 24 probability functions ${\mathbf
    P}_4(d)$ corresponding to the 24 permutations of the charge values
  given in table (\ref{corres}) using: (a) \emph{Drosophila}'s
  protein-coding sequence and (b) \emph{Drosophila}'s non-coding
  sequence (intergenic). Note that the shape of the probability function
  is invariant under the permutations.}
\label{fig8}
\end{figure}

\vskip 0.3in

\begin{figure}
\psfig{file=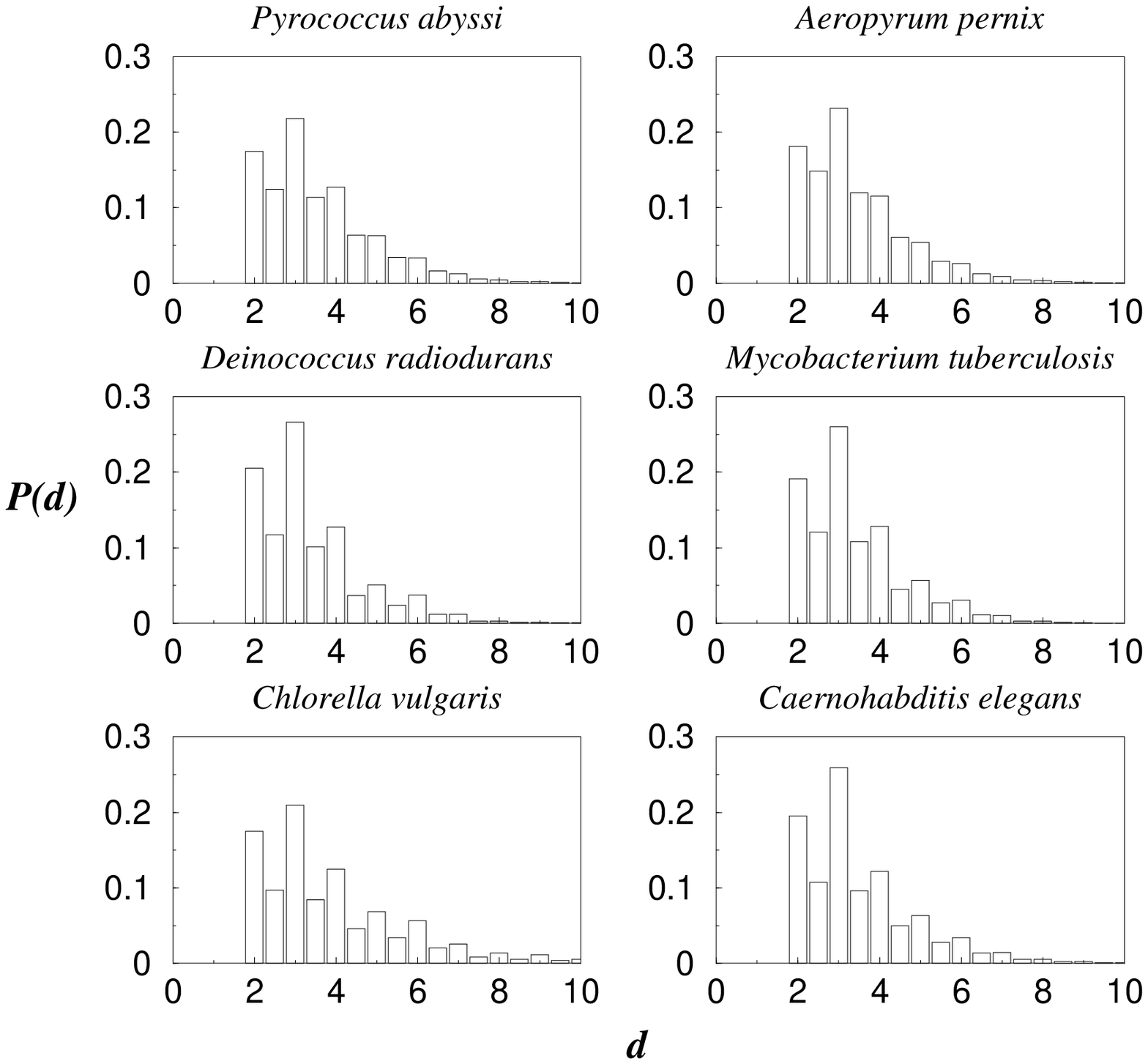,width=\hsize,clip=}
\caption[]{ Probability functions obtained when protein-coding
  sequences of different organisms are used for the assignment of
  monomer charges in the polymer. Note that in all the cases, the
  probability function reaches its maximum value at $d^*=3$. The above
  seems to be a generic property of protein-coding sequences.  }
\label{fig9}
\end{figure}

The ``peaking at $d^*=3$'' of the probability function ${\mathbf
  P}_4(d)$ seems to be a general characteristic associated with the
protein-coding sequences of living organisms, not only with
\emph{Drosophila}. In Fig.9 we show the probability functions
obtained from protein-coding sequences of different organisms, and
in all the cases the probability functions present their highest
value at $d^*=3$ (the mean distance $\bar{d}$ is also very close
to $3$).  The fact that the above characteristic is absent in
non-coding genetic sequences may be interpreted in evolutionary
terms. Genetic sequences directly involved in the
protein-translation processes were selected (among other things)
as to bring the distance between consecutive potential minima
closer to 3, both in mean and frequency of occurrence. This
interpretation raises a question: how likely is it to obtain a
randomly generated sequence with a structure similar to that of
protein-coding sequences?  In other words, if we generate a random
sequence and compute its probability function ${\mathbf P}_m(d)$,
how likely is it to come up with a probability function peaking at
$d^*=3$?.

In order to answer this question, we define the parameters $p_{3/2}$
and $p_{3/4}$ as

\vskip 0.3in

\begin{figure}
\psfig{file=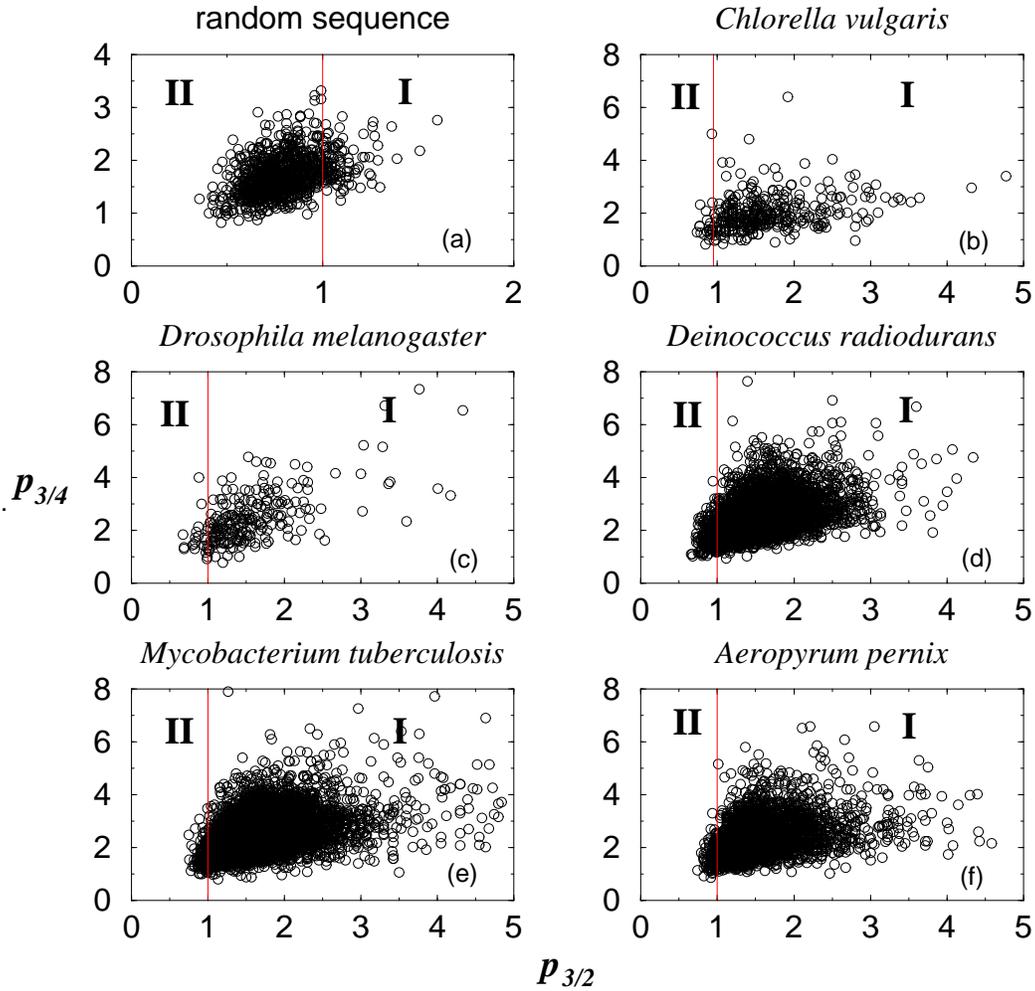,width=\hsize,clip=}
\caption[]{ Fluctuations of the probability function ${\mathbf
    P}_4(d)$ exhibited through the parameters $p_{3/2}$ and $p_{3/4}$.
  (a) Plot of $p_{3/2}$ versus $p_{3/4}$ for 4500 random sequences,
  each one 500 bases in length.  (b)-(f) The same as above but using
  protein-coding sequences of different organisms. The number of
  points vary in each graph since the available genetic sequences used
  in the calculations had different lengths. These sequences were
  divided into small pieces, each one composed of 500 bases. Note that
  for protein-coding sequences, the majority of the points fall in
  region I, for which both $p_{3/2}$ and $p_{3/4}$ are greater than 1.
  Note also the scale on the axes.  }
\label{fig10}
\end{figure}

\begin{eqnarray}
p_{3/2}&=&\frac{{\mathbf P}_4(3)}{{\mathbf P}_4(2)}\nonumber\\
\label{p32}\\
p_{3/4}&=&\frac{{\mathbf P}_4(3)}{{\mathbf P}_4(4)}\nonumber
\end{eqnarray}

If the probability function ${\mathbf P}_4(d)$ associated with a given
sequence has its highest value at $d^*=3$, then the corresponding
parameters $p_{3/2}$ and $p_{3/4}$ will both be greater than one.
Otherwise, one or both of these parameters will be smaller than one.

Fig.10a is a plot of $p_{3/2}$ vs. $p_{3/4}$ for 1000 random
sequences, each one consisting of 500 bases (which is a typical length
of sequences coding functional proteins \cite{genes}). It can be seen
from the figure that only a small fraction of the points (about 0.224)
fall in region I, for which $p_{3/2}>1$ and $p_{3/4}>1$. The rest of
the points fall in region II, in which $p_{3/2}<1$ and $p_{3/4}>1$.
Therefore, the probability of having a random sequence, 500 bases
long, whose consecutive potential minima are more often separated by a
distance $d^*=3$, is close to $0.224$.

On the other hand, Figs.10b-f show similar graphs, but using
protein-coding sequences of real organisms. These graphs were
constructed by analyzing short coding sequences 500 bases in length.
The fraction of points falling in region I ($p_{3/2}>1$ and
$p_{3/4}>1$) for the different organisms of Fig.10 is summarized in
the following table:

\begin{center}
\begin{tabular}{l|c}
\bf{Organism} & \bf{Frac. region I} \\ \hline
random sequence & 0.224000 \\
\emph{Chlorella vulg.} & 0.788235 \\
\emph{Drosophila} & 0.780220 \\
\emph{Deinococcus rad.} & 0.803525 \\
\emph{Myc. tuberculosis} & 0.802643 \\
\emph{Aeropyrum pernix} & 0.719973
\end{tabular}
\end{center}

These results show that protein-coding sequences are much more
likely to have their consecutive potential minima separated by a
distance $d^*=3$.

To end this section, we should make a further comment. The fact that
the most probable distance $d^*$ between consecutive potential minima
is $d^*=3$ for protein-coding sequences, is a consequence of the order
in which the bases or codons appear along the sequence and are not
related to the relative weight (fraction) of their occurrence. For
example, Fig.11a shows the probability function ${\mathbf P}_4(d)$
corresponding to a protein-coding sequence of \emph{Escherichia coli},
45000 bases in length. In Fig.11b the codon composition, expressed as
fractional occurrence, of this sequence is depicted. To construct this
last graph, we labelled each codon in an arbitrary way, by assigning
an integer number, from 0 for AAA up to 63 for GGG. On the horizontal
axis the number of the codon is plotted, and on the vertical axis its
fractional occurrence on the \emph{E. coli} sequence. Finally, Fig.11c
shows the probability function ${\mathbf P}_4(d)$ corresponding to a
random sequence, of the same length and codon composition as the one
used in Fig.11a. As can be seen, for the ``randomized'' \emph{E. coli}
sequence the most probable distance is no longer $d^*=3$ as in
Fig.11a, but $d^*=2$.

\begin{figure}
\psfig{file=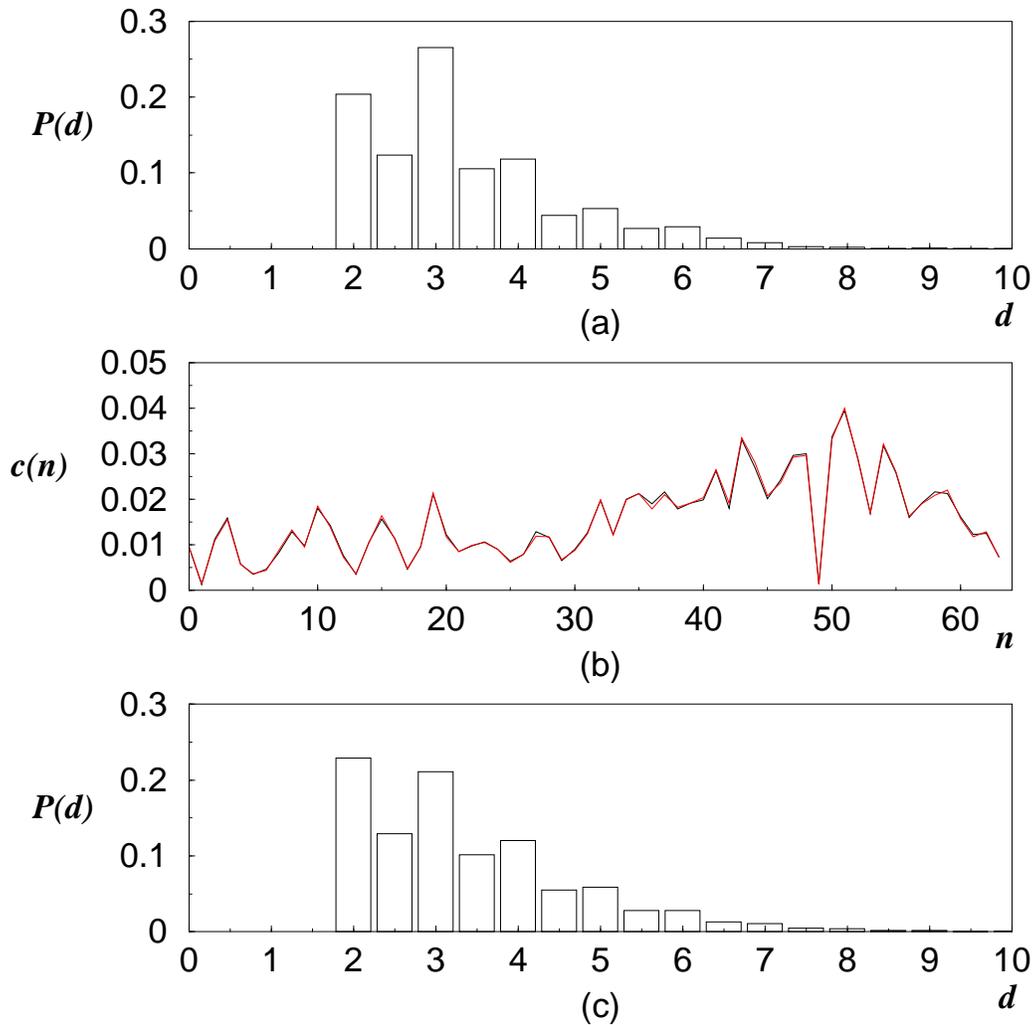,width=\hsize,clip=}
\caption[]{(a) Probability function corresponding to a
  protein-coding sequence of \emph{Escherichia coli}. (b) Codon
  composition $c(n)$ of this \emph{E. coli} sequence. Each codon was
  labeled with an integer number $n$, from 0 for AAA up to 63 for GGG.
  The horizontal axis is the codon number $n$ and the vertical axis is
  the codon composition $c(n)$ expressed as the fractional occurrence
  of codon $n$. (c) Probability function corresponding to a random
  sequence with the same codon composition as in (b). Note that for
  this randomized sequence, the peak at $d=3$ is lost.  }
\label{fig11}
\end{figure}


\section{Extended chain: $M>1$}
\label{polychain}

In this section we consider the case in which the chain is composed of
more than one single monomer. As mentioned before, this reflects the
fact that the ribosome is not a point particle after all. At each
given time, the ribosome interacts with the mRNA at several points,
giving rise to a collective interaction.  Electron microscopy has
revealed that the mRNA thread passes across the ribosome throughout a
\emph{tunnel} about 10nm long and 2nm in diameter \cite{genes}. Since
nucleotide length is about 0.5nm, around 20 nucleotides may be in a
position to interact simultaneously with the ribosome. Taking the
above into account, we will work with a small chain, assuming $M=10$
as a reasonable length.

In the step-like limit, the interaction potential equation can be
written by equation (\ref{Vcp}) now with the variables $\{V_j\}$
given by:

\begin{equation}
V_j=Kq_j\sum_{i=1}^Mp_i
\label{pcor}
\end{equation}

The main modification caused by using equation (\ref{pcor}) instead of
(\ref{Vparticle}) for the random variables $\{V_j\}$ is that they are
no longer statistically independent. Due to the collective interaction
between the chain and the polymer, these variables are strongly
correlated. This makes the problem of finding the probability function
${\mathbf P}_m(d)$ too difficult for analytical treatment. Therefore,
we will present only numerical results.

\begin{figure}
\psfig{file=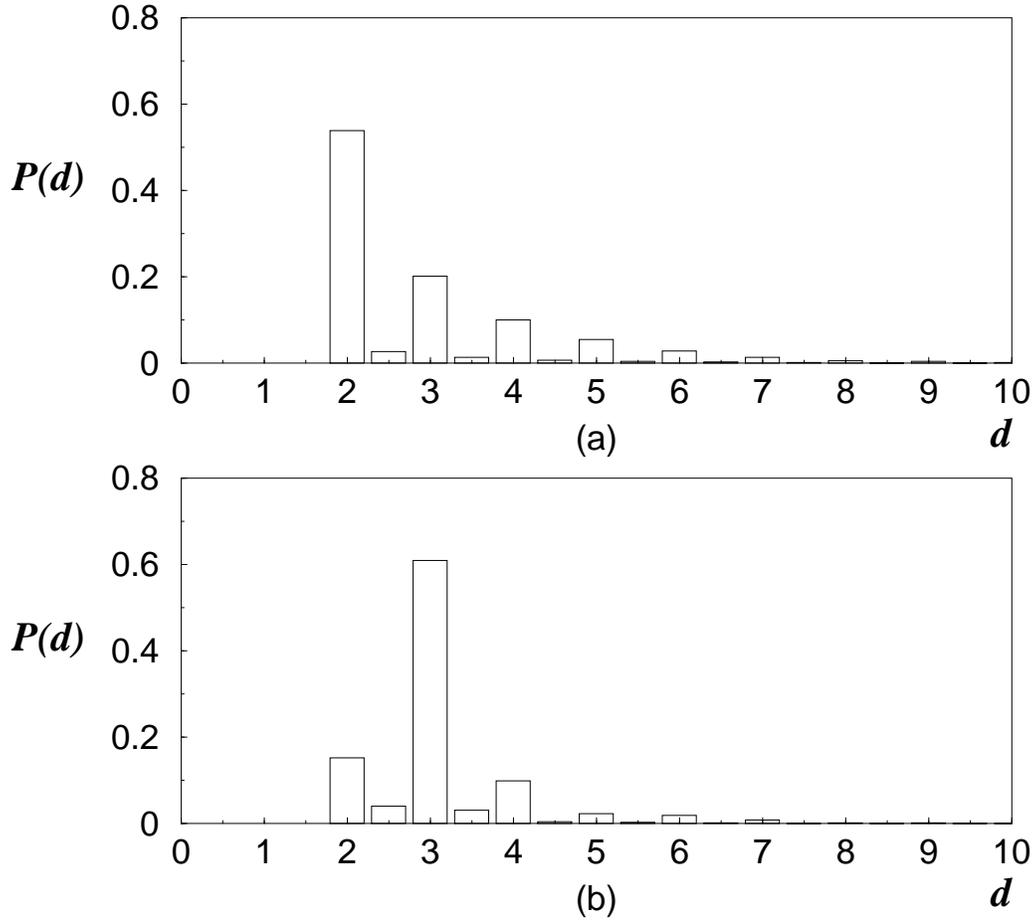,width=\hsize,clip=}
\caption[]{ Two graphs of the probability function ${\mathbf
    P}_4(d)$ corresponding to two different realizations of monomer
  charges for a 10-monomer chain interacting with a 500 monomer
  polymer. As can be seen, the shape of the probability function
  depends strongly on the particular realization of charges in the
  chain. The mean distance $\bar{d}$ is $\bar{d}\simeq 2.36$ for (a)
  and $\bar{d}\simeq 2.98$ for (b).}
\label{fig12}
\end{figure}

In our simulations, ${\mathbf P}_m(d)$ presents an erratic behavior.
The shape of this function strongly depends on the particular
realization of monomer charges in the chain. For example, Fig.12a
shows the probability function ${\mathbf P}_m(d)$ for a polymer 500
monomers in length and a particular realization of charges in a
10-monomer chain. The charge values we used to generate these graphs
were $\{\pm 1, \pm 2\}$. Fig.12b shows a similar graph but for a
different realization of charges along the chain (the realization of
charges in the polymer was the same in both cases). As can be seen,
these two graphs differ considerably: the first presents a very sharp
maximum at $d^*=2$, whereas the second does so at $d^*=3$. Wide
fluctuations in the probability function ${\mathbf P}_m(d)$ were
always present in our numerical simulations.

This fluctuating behavior did not occur in the single-monomer-chain
case.  It is apparent from Fig.10a that the fluctuations of the
probability function ${\mathbf P}_m(d)$ in that case are rather small:
the probability function has its highest value at $d^*=2$ for the
majority of the realizations (77.6$\%$) in the case $M=1$. Also, the
fluctuations are concentrated in a small region of the
$(p_{3/2},p_{3/4})$ plane.

\begin{figure}
\psfig{file=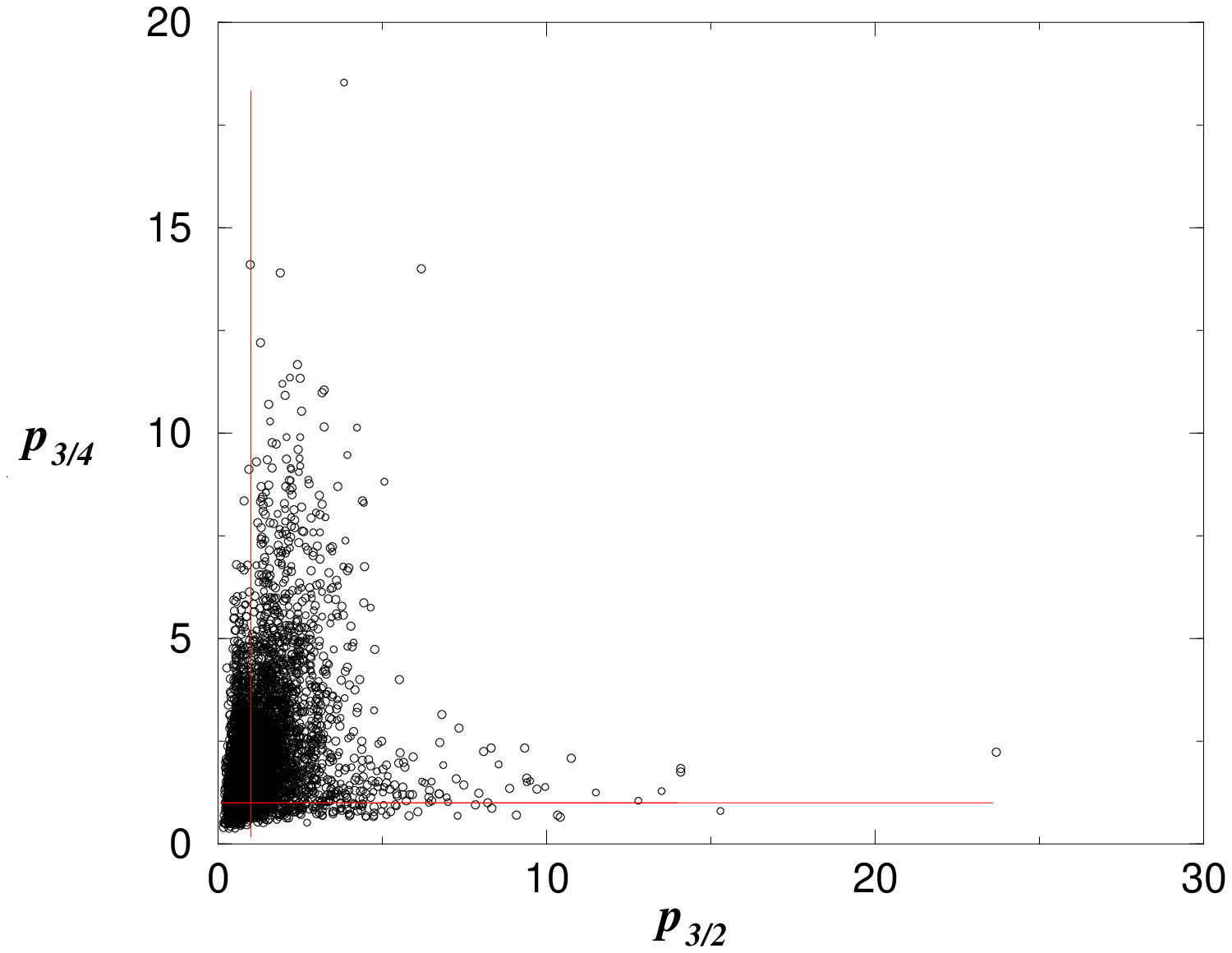,width=\hsize,clip=}
\caption[]{ Fluctuations of the probability function exhibited
  by the parameters $p_{3/2}$ and $p_{3/4}$ for the extended-chain case.
  A 10-particle chain was used along with a polymer 500 monomers in
  extent. The graph shows the above parameters for 4500 realizations
  or monomer charges in the chain. Note that the points spread out
  over a much wider region than in Fig.10a. Also, in this case the
  majority of the points fall in the region for which $p_{3/2}>1$ and
  $p_{3/4}>1$. }
\label{fig13}
\end{figure}

In order to have an idea of the magnitude of the fluctuations of the
probability function ${\mathbf P}_m(d)$ in the extended-chain case, we
can make use of the parameters $p_{3/2}$ and $p_{3/4}$ defined in
expression (\ref{p32}).  The fluctuations of the probability function
in the extenden-chain case are shown in Fig.13.  To construct this
figure, a polymer 500 monomers in length was used along with a
10-monomer chain.  The charge values in the polymer as well as in the
chain were again $\{\pm 1, \pm 2\}$. The parameters $p_{3/2}$ and
$p_{3/4}$ were calculated for 4500 charge realizations in the chain
(the realization of charges in the polymer was the same in all the
cases). Fig.13 is the plot of $p_{3/2}$ versus $p_{3/4}$ for these
4500 realizations.  Two remarks are worth mentioning about this
figure:

\begin{itemize}
\item The points are spread out over a much wider region than in the
  single-monomer-chain case (Fig.10a).
\item The majority of the points fall in the region in which both
  $p_{3/2}>1$ and $p_{3/4}>1$.
\end{itemize}

The following table shows the fractions of the points falling in each
of the four regions of the graph:

\begin{center}
\begin{tabular}{c|c}
\bf{Region} & \bf{Fraction} \\ \hline
$p_{3/2}>1$, $p_{3/4}>1$ & 0.52978 \\
$p_{3/2}<1$, $p_{3/4}>1$ & 0.33800 \\
$p_{3/2}>1$, $p_{3/4}<1$ & 0.07778 \\
$p_{3/2}<1$, $p_{3/4}<1$ & 0.05444
\end{tabular}
\end{center}

Therefore, when a collective interaction between the polymer and the
chain prevails, a remarkable property arises: The probability of
having a random interaction potential, whose consecutive minima are
more often separated by three monomers, is the largest.


\section{Dynamics}
\label{dynamics}

Recent experimental evidence suggests that the ribosome-mRNA system
presents a ratchet-like behavior in the protein synthesis
translocation process \cite{nature}.  In this view, the ribosome is
tightly attached to the mRNA thread in the absence of GTP. This is so
because the channel in the ribosome through which the mRNA passes, is
more or less closed.  When a GTP molecule is supplied (and transformed
into GDP), this channel opens leaving the mRNA thread free to move one
codon. Subsequently the mRNA passage in the ribosome closes again,
trapping the mRNA molecule.

In this clamping mechanism several physicochemical factors are
involved, which if taken into account in detail would lead to complex
dynamical equations hard to handle. In this work our approach is to
look into the behavior of oversimplified molecular models which might
capture some of the essential dynamical features of the system and may
shed some light on how this mechanism could have arisen in the origin
of life conditions.

In our modelling the dynamics of the system is governed by the
application of an external force $F_{ex}$ to the chain in the
horizontal direction, i.e. parallel to the polymer. By this means the
chain will be forced to move along the polymer. In principle, the
force $F_{ex}$ may be time dependent, but we will restrict ourselves
to a constant term. This force might come from a chemical pump (like
GDP) or from any other electromagnetic force present in prebiotic
conditions. The only purpose of this force in our model is to drive
the chain along the polymer (which is assumed to be fixed,
$N\rightarrow\infty$ limit), avoiding it from getting trapped in some
of the minima of the polymer-chain interaction potential $V(x)$.
Therefore, we will also assume that $F_{ex}$ satisfies
$|F_{ex}|>$ max$|\partial V(x)/\partial x|$.

Our analysis relies on Newton's equation of motion in a high friction
regime, where inertial effects can be neglected.  This regime actually
exists in biological molecular ratchets similar to the one we are
considering \cite{ratchet1,ratchet2}. Under such conditions, the
Newton's equation of motion acquires the form

\begin{equation}
\gamma\frac{dx}{dt}=-\frac{\partial V_{\sigma}^{\alpha}(x)}{\partial x}
+F_{ex}
\label{newton}
\end{equation}
where $\gamma$ is the friction coefficient. In what follows, we will
set $\gamma=1$, which is equivalent to setting the measure of the time
unit. The above, though a deterministic equation, gives rise to a
random dynamics due to the randomness of the interaction potential
$V_{\sigma}^{\alpha}(x)$. In order to start analyzing this random
dynamics, let us consider first the single-monomer-chain case $M=1$.
In this case, as before, we will refer to the chain simply as ``the
particle''.

\subsection{Single-monomer chain: $M=1$. Random sequences.}

\begin{figure}
\psfig{file=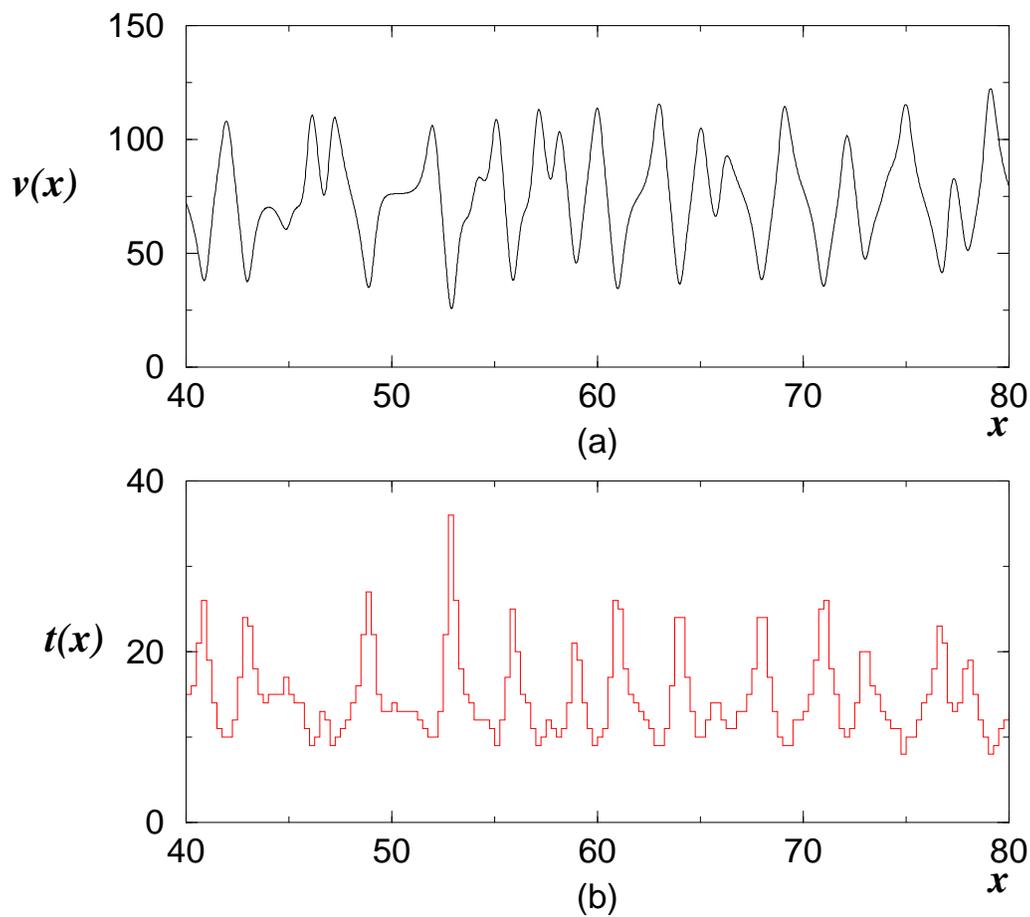,width=\hsize,clip=}
\caption[]{  (a) Typical realization of the velocity of the
  particle along the polymer as a function of the position $x$. (b)
  Local transit time $t$ as a function of the position $x$. Note that
  the particle effectively spends more time in certain regions of the
  polymer than in others.  Also, note the regularity in the positions
  of the maxima of the local transit time. }
\label{fig14}
\end{figure}

In Fig.14a we show a typical realization of the velocity of the
particle $v(x)$ as a function of its position $x$ along the polymer.
This graph was constructed by solving numerically the equation of
motion (\ref{newton}), using the fourth order Runge-Kuta method.  The
parameter values used were $\alpha=2$ and $\sigma=0.5$, and the
monomer charge values were $\{\pm 1,\pm 2\}$ (case $m=4$).  Fig.14b
shows the local-transit times of the particle along a short segment of
the polymer (40 monomers in length). This transit time is represented
in arbitrary units, and was computed by counting how many time steps
the particle spent in every spatial interval $\Delta x$ throughout the
polymer. In the graph shown, the value of $\Delta x$ was $\Delta
x=0.25$. It is apparent from this figure that, in its way along the
polymer, the particle spends more time in certain regions than in
others, the former being more or less regularly spaced along the
polymer.

In order to find out the spatial regularities in the dynamics of the
system, it is convenient to take the Fourier transform of the velocity
$v(x)$ of the particle. Let us call $\hat{v}(k)$ the Fourier transform
of $v(x)$, $k$ being the Fourier variable conjugate to $x$.

\begin{figure}
\psfig{file=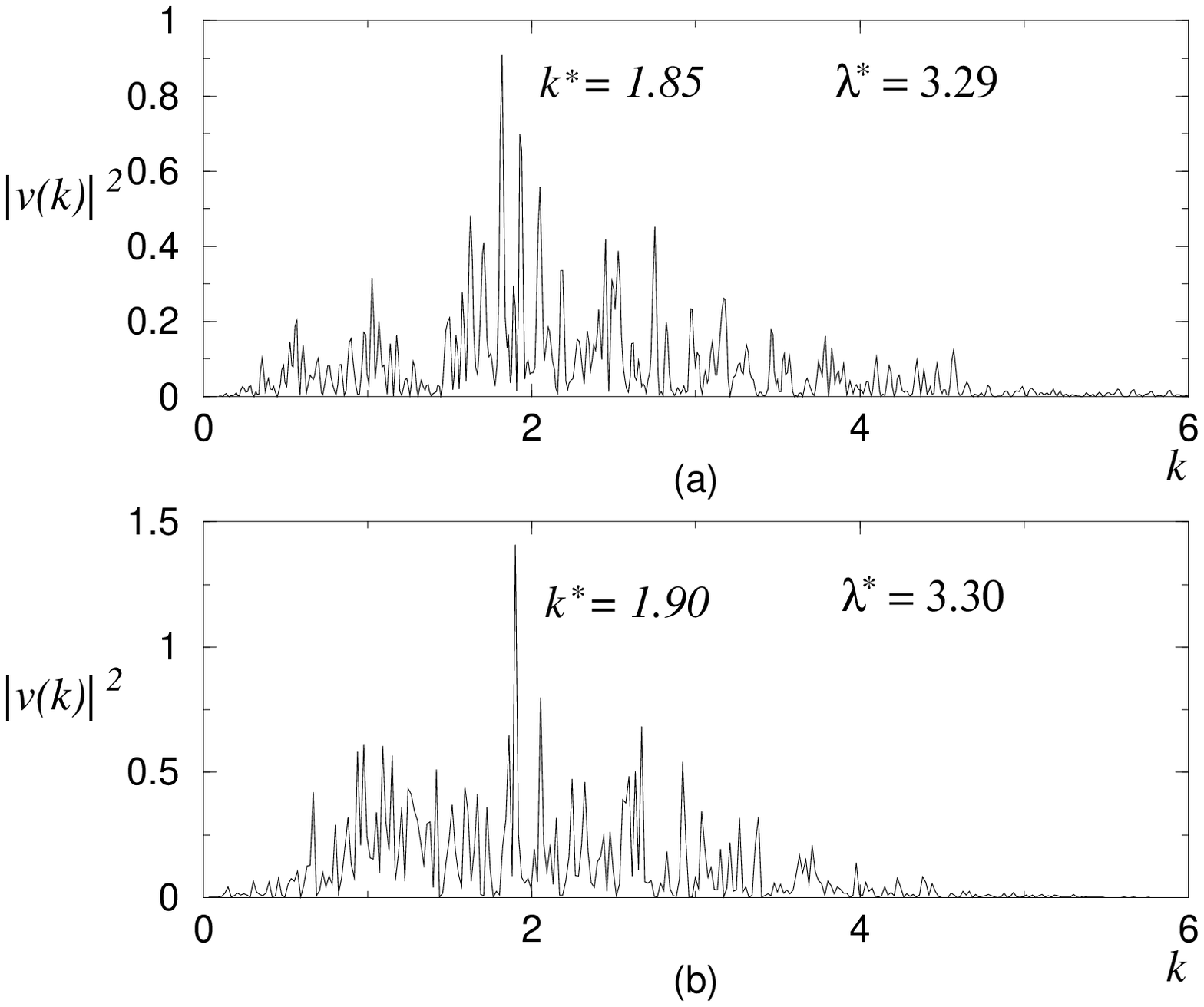,width=\hsize,clip=}
\caption[]{ Power spectrum of the velocity of the particle
  along the polymer, for two random realizations of monomer
  charges in the polymer.  Note that there exist dominant frequencies
  $k^*$ even though the charges of the polymer were assigned at
  random. These dominant frequencies correspond to spatial
  regularities $\lambda^*=2\pi/k^*$ with values (a)
  $\lambda^*\simeq 3.29$ and (b) $\lambda^*\simeq 3.30$.  }
\label{fig15}
\end{figure}

Fig.15 shows the Fourier power spectrum of the velocity,
$|\hat{v}(k)|^2$, for two different realizations of monomer charges in
the polymer. The parameter values in Fig.15a and Fig.15b are
$\{\sigma=0.5,\ \alpha=1\}$ and $\{\sigma=0.5,\ \alpha=4\}$
respectively. These graphs were computed for the case $m=4$, using the
charge values $\{\pm 1, \pm 2\}$. From the figure, it is evident that
there exists a dominant frequency $k^*$ in the power spectrum of the
velocity (the highest peak), whose corresponding spatial periodicity
is $\lambda^*=2\pi/k^*\sim3.3$.  The power spectrum reveals a
dynamical regularity in the motion of the particle throughout the
polymer. This regularity is inherited from the one present in the
random potential, in the sense that the particle spends more time in
the minima than in the maxima.  The consequence is a slowing down of
the velocity nearly every three monomers, which is reflected in the
power spectrum. Our interpretation is that the peak occurring in the
power spectrum of the velocity conveys the information on the average
distance $\bar{d}$ between consecutive potential minima, which for the
case $m=4$ is $\bar{d}\sim3.4$.

\subsection{Single-monomer chain: $M=1$. Real sequences.}

As in section \ref{realseq}, we can assign the charges along the
polymer in correspondence with the genetic sequence of real organisms.
In order to do that, we will use the same base-charge correspondence
as in expression (\ref{corres}). The objective is to find out how the
dynamics of the system changes when using real genetic sequences
instead of random ones. In what follows, the value of the parameters
$\alpha$ and $\sigma$ will be $\alpha=1$, $\sigma=0.5$.

\begin{figure}
\psfig{file=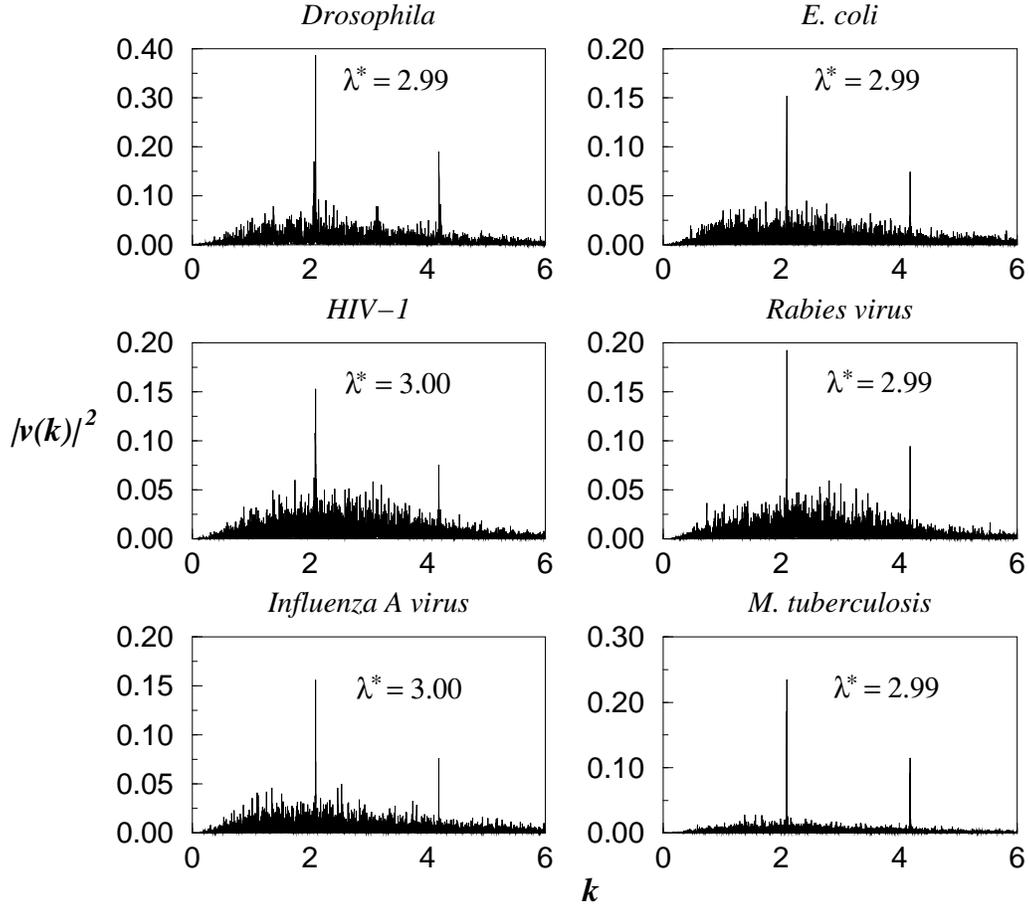,width=\hsize,clip=}
\caption[]{ Power spectrum of the velocity of the particle
  using protein-coding sequences of different organisms to assign the
  charges on the polymer. Note the very sharp peaks in all the
  spectra.  The above means that the dynamics generated by
  protein-coding sequences presents very well defined periodicities.
  Also, note that these spatial periodicities are almost equal to 3.
  The peak around $k=4$ is a resonant frequency (second harmonic) of
  the first peak. }
\label{fig16}
\end{figure}

\begin{figure}
\psfig{file=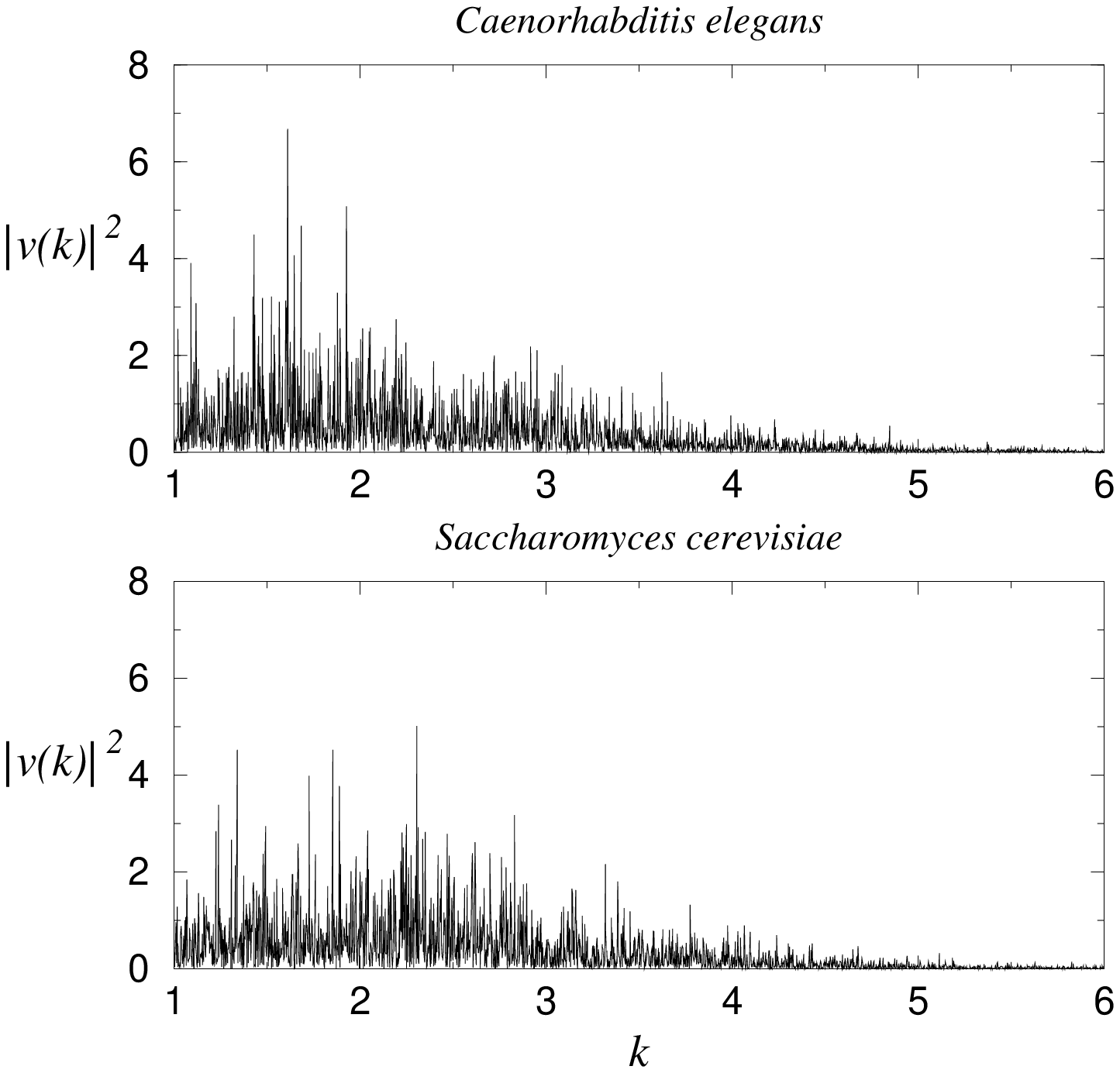,width=\hsize,clip=}
\caption[]{ Power spectrum of the velocity of the particle using
  non-coding sequences of \emph{S. cerevisiade} (yeast) and \emph{C.
    elegans} (worm). In this case the spatial periodicity is weaker
  (or absent) than in Fig.16.  It seems that the non-coding sequences
  of real organisms have a random like structure. }
\label{fig17}
\end{figure}

The power spectrum of the velocity of the particle throughout the
polymer, when using protein-coding sequences of different organisms,
is shown in Fig.16. To generate these graphs, short coding sequences
of several organisms, each 10000 monomers in length, were used. Two
points are worth noticing in this figure.  First, the peak in the
power spectrum is much higher than in the random case. This indicates
that there is a much more well defined periodicity in the dynamics
generated by interaction potentials when protein-coding sequences are
used. Second, the spatial periodicity reflected in the peak of the
power spectrum is much closer to $3$ than in the random case.

This dynamical behavior is not present when real but non-coding
sequences are used. For example, in Fig.17 we show the power spectrum
of the velocity of the particle along the polymer, for two cases in
which the monomer charges in the polymer were assigned in
correspondence with intergenic regions of two organisms.  As can be
seen, the structure of such spectra is similar to the one obtained in
the random case.  In this sense, intergenic regions again seem to have
a random structure.

The fact that the power spectrum corresponding to protein-coding
sequences exhibits a very sharp periodicity at $\lambda^*\simeq 3$,
whereas the one corresponding to non-coding sequences does not, has
already been reported in the literature \cite{lobzin}. Nonetheless, in
these previous works the power spectrum of the ``bare'' genetic
sequences is analyzed, namely, without considering any kind of
interaction potential or dynamical behavior. What we have shown here,
though, is that this ``structural'' periodicity around three
transforms into a \emph{dynamical periodicity} in the motion of the
particle along the polymer.

\subsection{Extended chain: M=10}

The most interesting dynamics occurs when an extended chain is
interacting with the polymer. In such a situation, a collective
interaction prevails. At every moment there are several contact points
between the chain and the polymer. As we have already pointed out,
collective interaction between the chain and the polymer gives rise to
a widely fluctuating probability function ${\mathbf P}_m(d)$. The same
occurs with the power spectrum of the velocity of the chain along the
polymer. However, these fluctuations, far from being annoying, produce
a much richer dynamical behavior than in the single-monomer-chain
case.

In Fig.18 we show the power spectra of the velocity of the chain along
the polymer for two different random realizations of monomer charges
in the chain. The charges in the polymer were the same in both cases.
These graphs were constructed with a polymer 500 monomers in length
and a 10-monomer chain. The parameter values used were $\sigma=0.5$
and $\alpha=1$. Also, the charge values were, as above $\{\pm 1,\ \pm
2\}$. From the figure, it is apparent that the power spectrum of the
velocity exhibits a very well defined dominant frequency, \emph{even
  though the charges in both the polymer and the chain were assigned
  at random}. The power spectrum in Fig.18a presents a dominant
frequency corresponding to a spatial periodicity $\lambda^*\simeq 2$,
whereas the corresponding periodicity in the power spectrum shown in
Fig.18b is $\lambda^*\simeq 3$. We should explain this difference.

\begin{figure}
\psfig{file=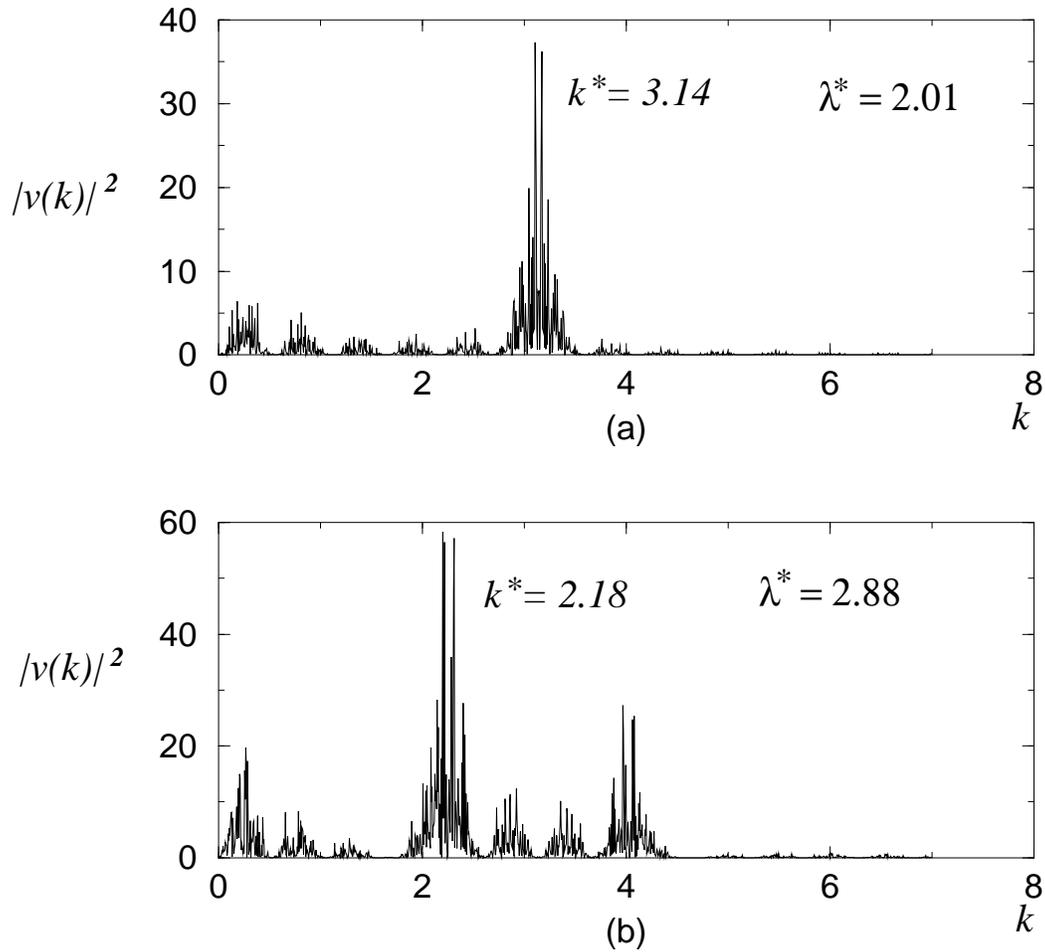,width=\hsize,clip=}
\caption[]{  Power spectrum of the velocity of a 10-monomer chain
  along a polymer 500 monomers in length. These two graphs correspond
  to two different random realizations of monomer charges in the
  chain. (a) The spatial periodicity is $\lambda^*\simeq 2.01$ and (b)
  $\lambda^*\simeq 2.88$.  In (a) the corresponding probability
  function had a shape like the one in Fig.12a with $d^*=2$ and
  $bar{d}\simeq 2$, whereas in (b) the probability function was as in
  Fig.12b with $d^*=3$ and $\bar{d}\simeq 3$. In the extended-chain
  case, there is a finite correlation lengt along the potential (equal
  to the size of the chain), which is reflected in the little
  ``heaps'' appearing in the power spectrum.}
\label{fig18}
\end{figure}

The power spectrum appearing in Fig.18a was constructed by using a
polymer and a chain whose associated probability function ${\mathbf
  P}_4(d)$ has the same shape as the one of Fig.12a. Namely, for this
system the probability function has a very high value at $d^*=2$.  On
the other hand, the power spectrum in Fig.18b corresponds to a system
whose probability function has a very sharp peak at $d^*=3$, as the
one shown in Fig.12b. From our numerical simulations, we can conclude
that \emph{whenever the probability function has a sharp maximum at a
  distance $d^*$, the power spectrum of the velocity also presents a
  sharp peak corresponding to a spatial periodicity $\lambda^*=d^*$.}

As we have seen, in the collective-interaction case the most probable
configurations are those in which the probability function ${\mathbf
  P}_4(d)$ has its highest value at $d^*=3$ (see Fig.13).  Therefore,
if we assign at random the monomer charges in the polymer and in the
chain, with high probability we will come up with a dynamics
possessing a very well defined periodicity: the chain will move along
the polymer in ``jumps'' whose length is nearly three monomers.


\section{Summary and discussion}
\label{summary}

The results presented throughout this work suggest a possible scenario
for the origin of the three base codon structure of the genetic code.
In this scenario, primitive one dimensional molecular machines,
initially with a random structure, exhibited a regular dynamics with a
``preference'' for a movement in steps of three bases.  By ``steps''
we mean a slowing down of the velocity of the chain along the polymer,
nearly every three monomers (see Fig.14b). Even in the simplest case
in which the chain consists of only one monomer, the above dynamical
regularity is apparent. We can think of the dynamics of primitive
molecular machines as being ``biased towards three''.

The preceding property is quite robust inasmuch as it hardly depends
on the particular kind of interaction between the polymer and the
chain.  On one hand, the kind of electrostatic potentials we have used
is representative of the actual interaction potentials between
particles occurring in Nature. These potentials are characterized in
our model by the parameter $\alpha$. We have also seen that the
distribution distances between neighboring maxima and minima along the
interaction potential, characterized by the probability function
${\mathbf P}_m(d)$, does not depend on this parameter (for small
values of $\sigma$), i.e. it will be the same whether the interaction
is coulombian or dipolar or of any other (electrostatic) type.

On the other hand, the spatial distribution of interaction potential
minima also does not depend on the particular values of the monomer
charges $\{q_j\}$ and $\{p_i\}$, as long as these values are of the
same order of magnitude. An important feature that the charges must
comply with is that they take more than two different values. This
allows for an order relation to be established among the different
types of monomers, leading to a maxima and minima structure of the
interaction potential. The probability function ${\mathbf P}_m(d)$,
which gives the probability of two consecutive minima being separated
by a distance $d$, only depends on $m$, namely, on the number of
different types of monomers. As $m$ increases, the mean distance
$\bar{d}$ between consecutive potential minima approaches three.
Nevertheless, in the particle (single-monomer-chain) case, the most
probable distance is $d^*=2$ (for $m>2$).

Still in this particle case, considerable changes take place when the
charges along the polymer are assigned in correspondence with
protein-coding genetic sequences of real organisms. In this case, not
only is the mean distance $\bar{d}$ between neighboring potential
minima nearly three, but also the most probable one, $d^*$, happens to
be three. This is a remarkable property of protein-coding sequences,
perhaps acquired throughout evolution.  Furhtermore the fact that this
``refinement'' is absent in non-coding sequences of real organisms,
strongly suggests that it is a consequence of the dynamical processes
involved in the protein synthesis mechanisms.

This interpretation is supported by the results obtained when the
\emph{dynamics} of the particle moving along the polymer is
considered.  In the random sequence case, there are dominant
frequencies in the power spectrum of the particle velocity related
with the spatial regularities of the interaction potential.  Moreover
in the protein-coding sequence case, the power spectrum of the
velocity shows a very well defined periodicity corresponding almost
exactly to a spatial distance $\lambda^*=3$.  Again, this behavior
does not occur for non-coding sequences of real organisms, which are
not involved in the translation processes.

A richer dynamics emerges when the chain is composed of several
mo\-no\-mers.  In this, more realistic, collective-interaction case,
the probability function ${\mathbf P}_m(d)$ presents very wide
fluctuations, depending on the particular assignment of monomer
charges in the chain. Nevertheless, the most probable
configurations are those for which the probability function has
its highest value at $d^*=3$. For these configurations, the power
spectrum of the chain velocity along the polymer exhibits a very
well defined spatial periodicity at $\lambda^*\simeq 3$.

Our results suggest an origin of life scenario in which primordial
molecular machines of chains moving along polymers in quasi
one-dimensional geometries, that eventually led to the protein
synthesis processes, were biased towards a dynamics favoring the
motion in ``steps'' or ``jumps'' of three monomers. The higher
likelyhood of these primitive``ribosomes'' may have led to the
present ribosomal dynamics where mRNA moves along rRNA in a
channel conformed by the ribosome. Dynamics may have acted in this
sense as one of the evolutionary filter favoring the three base
codon composition of the genetic code.

\vskip 0.5 in
\noindent {\large \bf Acknowledgements}\\
\noindent We would like to thank Leo Kadanoff, Sue Coppersmith,
Haim Diamant and Cristian Huepe for very useful discussions and
corrections. This work was sponsored by the DGAPA-UNAM project
IN103300, the MRSEC Program of the National Science Foundation (NSF)
under Award Number DMR 9808595 and by the NSF Program DMR 0094569. M.
Aldana also acknowledges CONACyT-M\'exico a posdoctoral grant.

\pagebreak


\pagebreak


\end{document}